\documentclass[sigconf]{acmart}

\usepackage{graphicx}
\usepackage{siunitx}
\usepackage{enumitem}

\usepackage[acronym,shortcuts]{glossaries}
\newacronym{FOV}{FOV}{field of view}
\newacronym{HMD}{HMD}{head-mounted display}
\newacronym{MAFC}{MAFC}{multiple-alternative forced-choice}
\newacronym{OLS}{OLS}{ordinary least squares}
\newacronym{RP}{RP}{retinitis pigmentosa}
\newacronym{SPV}{SPV}{simulated prosthetic vision}
\newacronym{VPU}{VPU}{vision processing unit}
\newacronym{VR}{VR}{virtual reality}

\usepackage{xcolor}

\AtBeginDocument{%
  }

\setcopyright{acmlicensed}
\copyrightyear{2018}
\acmYear{2018}
\acmDOI{XXXXXXX.XXXXXXX}

\acmConference[Conference acronym 'XX]{Make sure to enter the correct
  conference title from your rights confirmation emai}{June 03--05,
  2018}{Woodstock, NY}
\acmISBN{978-1-4503-XXXX-X/18/06}

\begin{document}


\title{Improving Wayfinding in Immersive VR Simulations of Bionic Vision through Temporal Semantic Simplification}
\title{Effects of Semantic Simplification Timing on Wayfinding Performance in Immersive Simulated Bionic Vision}
\title{Static or Temporal? Semantic Scene Simplification to Aid Wayfinding in Immersive Simulations of Bionic Vision}


\author{Justin M. Kasowski}
\affiliation{
    \institution{University of California}
    \city{Santa Barbara}
    \state{CA}
    \country{USA}
}
\email{justin_kasowski@ucsb.edu}

\author{Apurv Varshney}
\affiliation{
    \institution{University of California}
    \city{Santa Barbara}
    \state{CA}
    \country{USA}
}
\email{apurv@ucsb.edu}

\author{Michael Beyeler}
\affiliation{
    \institution{University of California}
    \city{Santa Barbara}
    \state{CA}
    \country{USA}
}
\email{mbeyeler@ucsb.edu}

\renewcommand{\shortauthors}{Anonymous et al.}

\begin{abstract}
Visual neuroprostheses (\emph{bionic eyes}) aim to restore a rudimentary form of vision by translating camera input into patterns of electrical stimulation. 
To improve scene understanding under extreme resolution and bandwidth constraints, prior work has explored computer vision techniques such as semantic segmentation and depth estimation. However, presenting all task-relevant information simultaneously can overwhelm users in cluttered environments.
We compare two complementary approaches to semantic preprocessing in immersive virtual reality: \emph{SemanticEdges}, which highlights all relevant objects at once, and \emph{SemanticRaster}, which staggers object categories over time to reduce visual clutter. 
Using a biologically grounded simulation of prosthetic vision, 18 sighted participants performed a wayfinding task in a dynamic urban environment across three conditions: edge-based baseline (\emph{Control}), \emph{SemanticEdges}, and \emph{SemanticRaster}. 
Both semantic strategies improved performance and user experience relative to the baseline, with each offering distinct trade-offs: \emph{SemanticEdges} increased the odds of success, while \emph{SemanticRaster} boosted the likelihood of collision-free completions.
These findings underscore the value of adaptive semantic preprocessing for prosthetic vision and, more broadly, may inform the design of low-bandwidth visual interfaces in XR that must balance information density, task relevance, and perceptual clarity.
\end{abstract}


\begin{teaserfigure}
    \centering
    \includegraphics[width=\linewidth]{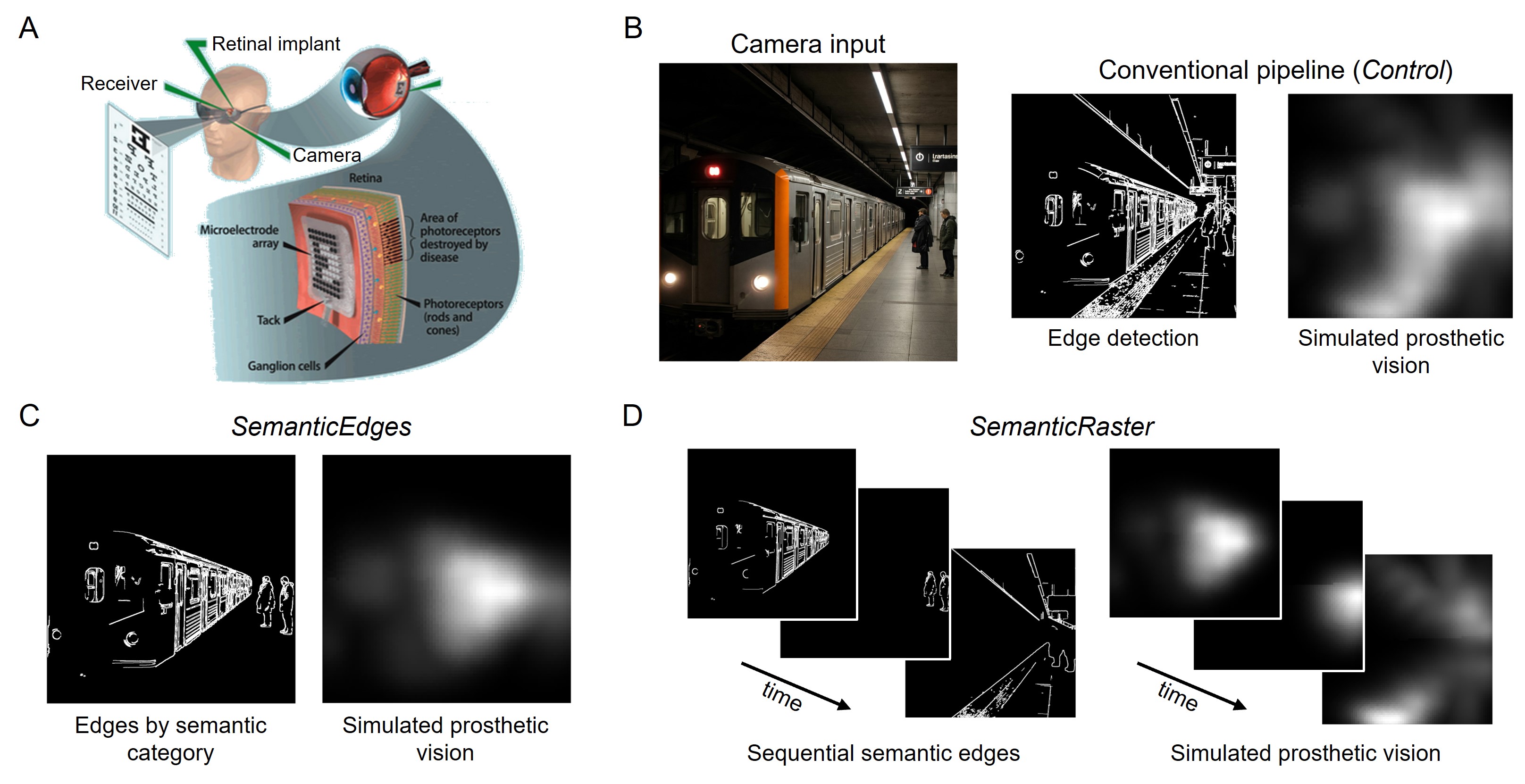}
    \caption{Scene simplification for bionic vision. 
    (A) Bionic vision systems (e.g., retinal implants) capture visual information via an external camera and deliver it to a microelectrode array implanted in the visual system. 
    (B) Traditional preprocessing methods, like edge detection (\emph{Control}), highlight basic scene features but fail to prioritize task-relevant information. 
    (C) \emph{SemanticEdges} enhances the scene by isolating key semantic groups (e.g., pedestrians, obstacles) while suppressing irrelevant background details.
    (D) The novel \emph{SemanticRaster} extends this approach by sequentially presenting semantic groups over time, prioritizing navigational hazards to reduce clutter and enhance scene understanding in dynamic environments.}
  \label{fig:teaser}
\end{teaserfigure}


\maketitle

\section{Introduction}

By 2050, over 114 million people are expected to be living with incurable blindness, representing a major global health challenge~\cite{bourne_global_2020}.
Electronic visual prostheses, or \emph{bionic eyes}, offer a promising solution for individuals with retinal degeneration, optic nerve damage, or cortical injury~\cite{weiland_electrical_2016,fernandez_development_2018}. These devices capture visual input from an external camera, process it via a vision processing unit (VPU), and deliver electrical stimulation to neurons in the retina, optic nerve, or visual cortex~\cite{luo_argusr_2016,palanker_photovoltaic_2020,fernandez_visual_2021,titchener_second-generation_2022,troyk_intracortical_2017}, producing percepts known as \emph{phosphenes} to support basic tasks such as navigation and object localization~\cite{titchener_second-generation_2022,geruschat_analysis_2016}.

Despite clinical advances, existing systems like the Argus II \citep{luo_argusr_2016} and Bionic Vision Australia's suprachoroidal implant \citep{titchener_second-generation_2022} provide only a coarse, pixelated view of the world (Fig.\ref{fig:teaser}B), limited by low electrode counts and narrow fields of view. Even as next-generation devices improve electrode density~\cite{palanker_photovoltaic_2020,chen_shape_2020,jung_stable_2024}, strict safety regulations constrain how many electrodes can be activated simultaneously, limiting the system's effective resolution.

To maximize the utility of this constrained stimulation channel, commercial devices like the Argus II activate only subsets of electrodes (\emph{timing groups}) in rapid temporal succession~\cite{second_sight_argus_2013}.
This raster-scanning approach, inspired by display technology, divides the visual field into strips that are activated in sequence to create the perception of a coherent image.
Raster patterns—defining the spatial layout and order of stimulation—are typically chosen heuristically and remain agnostic to scene content.
Recent work by \citet{kasowski_simulated_2025} showed that a checkerboard raster, which maximizes spatial distance between simultaneously active electrodes, improves clarity and task performance over vertical, horizontal, or random arrangements, while complying with safety limits.

Meanwhile, research has focused on preprocessing strategies that simplify visual input prior to stimulation.
Semantic segmentation~\cite{sanchez-garcia_semantic_2020,han_deep_2021} can isolate important object classes such as pedestrians, vehicles, or obstacles (Fig.\ref{fig:teaser}C), and depth-based strategies can emphasize near-field hazards\cite{mccarthy_mobility_2014,sadeghi_benefits_2024,rasla_relative_2022}.
However, even these simplified images often overwhelm the user when displayed all at once, especially under tight stimulation constraints that limit how many electrodes can be active in a given frame~\cite{beyeler_towards_2022}.

We propose a novel content-aware raster strategy called \emph{SemanticRaster}, which bridges these two perspectives.
Rather than activating spatial strips or checkerboards, the system cycles through semantic groups over time: for example, first displaying hazards like cars or bicycles, then pedestrians, then structural elements (Fig.\ref{fig:teaser}D).
This approach aims to reduce clutter and direct attention to task-relevant features while maintaining context across frames.
The prioritization of object categories is flexible and ideally co-designed with blind users\cite{reis_patient_2011,chung_large-scale_2024,nadolskis_aligning_2024}, offering a foundation for temporally adaptive encoding that reflects user needs and task demands.

Because no commercial retinal implants are currently available (and clinical testing is constrained by risk, device heterogeneity, and small sample sizes), direct evaluation of raster strategies in end users is infeasible. \Ac{SPV} in immersive \ac{VR} provides a powerful alternative~\cite{hayes_visually_2003,dagnelie_real_2007,kasowski_immersive_2022}, enabling precise, repeatable testing of design strategies in realistic settings.
We use \texttt{BionicVisionXR}\cite{kasowski_immersive_2022}, an open-source VR platform with gaze-contingent rendering and psychophysically grounded models of phosphene appearance\cite{beyeler_model_2019}, temporal dynamics~\cite{hou_predicting_2024}, and spatial summation~\cite{hou_axonal_2024}. These models simulate how a future epiretinal device would respond to head and eye movements in dynamic environments.

In this study, 18 sighted participants completed a wayfinding task through a cluttered virtual city using simulated prosthetic vision. While sighted participants cannot model long-term perceptual learning, they enable controlled, within-subject comparisons that are impractical in implant users~\cite{beyeler_learning_2017}. Gaze-contingent rendering allowed us to emulate the visual experience of a head-mounted camera system interacting with retinal stimulation.

Our work makes three key contributions:
\begin{enumerate}[topsep=0pt,itemsep=-1ex,partopsep=0pt,parsep=1ex,leftmargin=14pt,label=\roman*.]
    \item We introduce \emph{SemanticRaster}, a content-aware raster strategy that sequences semantic groups over time, offering a new method for reducing clutter while preserving context under tight stimulation constraints.
    \item We conduct a controlled user study in immersive VR that systematically compares static and temporally adaptive semantic encoding strategies using realistic phosphene simulations and dynamic obstacles.
    \item We show that static and temporally sequenced semantic simplification confer complementary benefits (higher completion rates and lower collision rates, respectively) providing design guidance for bandwidth‑limited XR and next‑generation bionic‑vision interfaces.
\end{enumerate}

\section{Background}

Visual neuroprostheses, or bionic vision systems, aim to restore rudimentary visual function to people with profound blindness by bypassing damaged visual pathways and directly stimulating surviving neurons~\cite{weiland_electrical_2016,fernandez_development_2018}. Depending on the site of implantation, these devices target the retina, optic nerve, or visual cortex.

Retinal implants such as the Argus II~\cite{luo_argusr_2016}, Alpha-IMS~\cite{stingl_subretinal_2015}, and suprachoroidal devices~\cite{titchener_second-generation_2022} represent the most clinically advanced prostheses to date. Meanwhile, next-generation systems, such as PRIMA~\cite{lorach_photovoltaic_2015,palanker_photovoltaic_2020},  ICVP~\cite{jung_stable_2024}, and Neuralink's cortical array~\cite{musk_integrated_2019}, seek to improve spatial resolution and usability through denser electrode layouts and more flexible implantation strategies.

Most of these systems rely on an external visual processing unit (VPU) to convert real-time video into stimulation patterns for the implanted electrode array. While electrical activation can elicit phosphenes (i.e., discrete points of light perceived by the user) the resulting vision is highly degraded: resolution is limited~\cite{wilke_electric_2011,fine_virtual_2024}, visual fields are narrow (e.g., $\sim10 \times 20$ deg in Argus II)\cite{thorn_virtual_2020}, and the appearance of phosphenes is variable and often distorted by biological factors\cite{beyeler_model_2019,sinclair_appearance_2016}. Safety limits on simultaneous electrode activation further constrain the effective resolution, even as newer devices push electrode counts into the hundreds~\cite{second_sight_argus_2013}.

As a result, users often describe prosthetic vision as unreliable, effortful, and situationally useful at best~\cite{nadolskis_aligning_2024}. Navigation and scene understanding remain especially challenging, as the limited field of view necessitates continuous head scanning to piece together a coherent sense of the environment~\cite{erickson-davis_what_2021}. Most current systems do not account for eye movements~\cite{caspi_eye_2021}, further complicating perceptual stability.

To improve usability and support greater independence, future prosthetic systems must not only improve hardware, but also intelligently preprocess visual input. This includes prioritizing task-relevant information, reducing clutter, and adapting to the user's context and behavior~\cite{beyeler_towards_2022}. Simulated prosthetic vision (SPV) in immersive VR has emerged as a powerful tool to prototype and evaluate such strategies, enabling rapid iteration without the need for implantable hardware.

\section{Related Work}

Bionic vision systems typically rely on preprocessing strategies to enhance usability within the constraints of low-resolution, pixelated visual input. Early approaches emphasized edge detection and contrast enhancement to make scene structure more perceptible~\cite{dagnelie_real_2007,vergnieux_simplification_2017}, though these methods often lacked adaptability to specific tasks or environments.

Recent work has explored more sophisticated computer vision methods, such as semantic segmentation and depth-based scene parsing, to prioritize task-relevant features like obstacles, pedestrians, or walkable paths~\cite{sanchez-garcia_semantic_2020,han_deep_2021}. For example, RGB-D-based preprocessing has been used to highlight nearby hazards~\cite{perez-yus_depth_2017,mccarthy_mobility_2014,sadeghi_benefits_2024}, while semantic approaches offer a schematic representation of the environment~\cite{sanchez-garcia_semantic_2020}. However, these methods often render all features simultaneously, leading to visual clutter, which is especially problematic under the perceptual constraints of prosthetic vision~\cite{han_deep_2021,lieby_substituting_2011}.

These limitations have motivated the search for dynamic prioritization strategies that balance clarity and informational value. In particular, time-multiplexed rendering strategies offer a promising way to sequence relevant scene elements, though few studies have explored this space systematically. Kasowski et al.~\cite{kasowski_simulated_2025} showed that temporally rasterized stimulation can improve visual decoding for simple tasks like letter identification. However, their work was limited to idealized, static stimuli, leaving open the question of whether similar benefits extend to more complex, dynamic environments.

Simulated prosthetic vision (SPV) in immersive virtual reality (VR) has emerged as a powerful testbed for evaluating encoding strategies prior to clinical deployment. These platforms allow sighted participants to act as ``virtual patients,'' experiencing key constraints of bionic vision (such as reduced resolution, limited field of view, phosphene blur, and temporal distortions) without the variability introduced by long-term perceptual adaptation or device-specific idiosyncrasies. While not a substitute for real-world testing, SPV enables controlled, repeatable, within-subject comparisons that are impractical in clinical studies, especially during early-stage prototyping~\cite{kasowski_immersive_2022,thorn_virtual_2021,rasla_relative_2022}. 
Early SPV studies relied on oversimplified visual models~\cite{dagnelie_real_2007}, but more recent work has introduced psychophysically validated phosphene simulations incorporating fading, spatial distortion, and gaze contingency~\cite{beyeler_model_2019,kasowski_immersive_2022}. Still, these advances have largely lacked temporally adaptive encoding aligned with users' moment-to-moment navigation goals.

Our work builds on this foundation by integrating (i) a biologically grounded, gaze-contingent phosphene simulation; (ii) semantically informed image processing; and (iii) a novel raster strategy that sequences object categories based on task relevance. Unlike prior approaches that treat semantic segmentation as static, \emph{SemanticRaster} encodes temporal prioritization to emphasize critical cues (e.g., moving obstacles) while minimizing clutter.

Though motivated by bionic vision, our framework may offer general-purpose strategies for temporally adaptive scene simplification in constrained visual displays. By combining perceptual realism, gaze contingency, and task-aware encoding, this work advances the design of real-time, user-centered interfaces for immersive and assistive technologies alike.

\section{Methods}

\subsection{Participants}

Eighteen participants with normal or corrected-to-normal vision (11 female, 7 male; ages 18–40; $M = 25.04$, $SD = 5.72$) were recruited for this study. 
Participants were recruited from the research participant pool at Anonymous University and served as ``virtual patients''~\cite{kasowski_immersive_2022} in \ac{SPV} experiments.

Prior experience with \ac{VR} varied: five participants had never used \ac{VR}, while the remaining 13 reported familiarity with the technology, ranging from 1 to over 20 prior sessions. To minimize risks of discomfort, participants with known sensitivity to flashing lights or motion sickness were excluded during the initial screening process.

The study adhered to the principles of the Declaration of Helsinki and was approved by the Institutional Review Board at Anonymous University.

\begin{figure*}[t!]
    \centering
    \includegraphics[width=\linewidth]{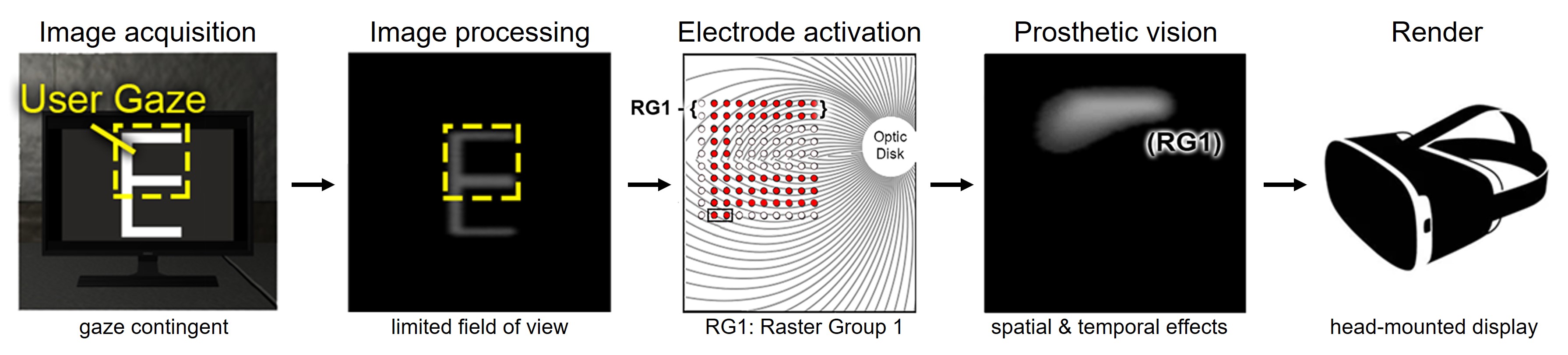}
    \caption{
        Simplified overview of the \ac{SPV} pipeline. 
        Unity's virtual camera captured scenes while tracking gaze position (``Image acquisition"). Frames underwent scene simplification, scaling, and grayscale conversion to mimic preprocessing by a visual processing unit (``Image processing"). Electrode activation levels were derived from pixel intensities, with temporal sequencing strategies grouping electrode activations over time (``Electrode activation"). The example illustrates grouped activation of the top two rows of electrodes. Spatial distortions were modeled using an axon map, and temporal effects like fading and persistence were integrated to simulate prosthetic vision (``Prosthetic vision"). The resulting percept was rendered to participants via a head-mounted display (``Render").}
    \label{fig:spv-pipeline}
\end{figure*}

\subsection{Simulated Prosthetic Vision}

We utilized the open-source Unity toolbox \texttt{BionicVisionXR} (\url{https://github.com/bionicvisionlab/BionicVisionXR}),  to simulate prosthetic vision within an immersive \ac{VR} environment. 
Participants viewed stimuli through an HTC VIVE Pro Eye head-mounted display, with phosphene appearance modeled using psychophysically validated simulations~\cite{horsager_predicting_2009,beyeler_model_2019,granley_computational_2021}. 
These simulations incorporated spatiotemporal dynamics, including phosphene elongation and fading due to axonal pathways~\cite{hou_axonal_2024} (Section~\ref{sec:spatial-model}), as well as persistence and decay effects based on charge accumulation dynamics~\cite{horsager_predicting_2009} (Section~\ref{sec:temporal-model}). 

To approximate the visual experiences of retinal prosthesis users, the \ac{VR} environment featured gaze-contingent rendering (Section \ref{sec:gaze}), dynamically updating scene content based on participants' head and eye movements. 
This ensured a realistic and interactive simulation of prosthetic vision.

We simulated a $10 \times 10$ epiretinal electrode array centered over the fovea, inspired by the Argus II implant~\cite{luo_argusr_2016}.
Electrodes were modeled as point sources with \SI{400}{\micro\meter} spacing, consistent with current-generation retinal prostheses. 
All simulations were rendered on a high-performance desktop computer (Intel i9-11900k, 64GB RAM, Nvidia RTX3090) and wirelessly transmitted to the head-mounted display.

This setup balances generalizability with alignment to near-future prosthetic technologies, providing a robust platform for evaluating visual preprocessing strategies in simulated prosthetic vision.
The entire \ac{SPV} workflow was thus as follows (Fig.~\ref{fig:spv-pipeline}):
\begin{enumerate}[topsep=0pt,itemsep=-1ex,partopsep=0pt,parsep=1ex,leftmargin=14pt,label=\roman*.]
    \item \textbf{Image acquisition:} Unity’s virtual camera captured a \SI{60}{\degree} field of view, rendered at \SI{90}{Hz}.
    \item \textbf{Image processing:} Frames were downscaled to $200 \times 200$ pixels, converted to grayscale, and smoothed with a $3 \times 3$ Gaussian kernel.
    \item \textbf{Electrode activation:} Pixel intensities nearest to each electrode were used to compute activation levels.
    \item \textbf{Spatiotemporal effects:} Phosphene shapes were modeled using the axon map model~\cite{beyeler_model_2019,granley_computational_2021}, simulating elongated phosphenes aligned with retinal ganglion cell axons. A temporal model~\cite{horsager_predicting_2009} simulated phosphene fading and persistence by accounting for charge accumulation and decay.
    \item \textbf{Gaze-contingent rendering:} The implant location dynamically shifted based on gaze position, ensuring the scene remained aligned with participants' fixation.
\end{enumerate}

\subsubsection{Spatial Distortions}
\label{sec:spatial-model}

The shape of phosphenes in epiretinal devices is influenced by the retinal ganglion cell axons, which traverse the retina in curved paths~\cite{rizzo_perceptual_2003,beyeler_model_2019}. 
We used the axon map model to simulate these distortions \citep{beyeler_model_2019,granley_computational_2021}.
Each electrode activated a region of the retina defined by Gaussian falloff parameters $\rho$ (spread) and $\lambda$ (elongation).
The instantaneous brightness $b_I$ of each pixel $(r,\theta)$ in the percept was computed according to:
\begin{equation}
    b_I = \max_{p \in R(\theta)} \sum_{e \in E} \exp \bigg( \frac{-d_e^2}{2\rho^2} + \frac{-d_\mathrm{soma}^2}{2\lambda^2} \bigg),
    \label{eq:axon-map}
\end{equation}
where $R(\theta)$ is the path of the axon terminating at retinal location $(r,\theta)$, $p$ is a point along the path, $d_e$ is the distance from $p$ to the stimulating electrode $e$, and $d_\mathrm{soma}$ is the distance along the axon from $p$ to the cell body. 
Spatial distortions were modeled using medium levels of elongation and spread, as reported in earlier psychophysical studies~\cite{beyeler_model_2019}, with $\rho=\SI{200}{\micro\meter}$ (spread) and $\lambda=\SI{400}{\micro\meter}$ (elongation). 
These parameters were selected to represent typical distortions experienced by prosthesis users, balancing realism and perceptual clarity for the purposes of the study. 
By keeping these parameters constant across conditions, we ensured that observed differences in performance were attributable to the preprocessing strategies rather than variations in spatial distortions.

\subsubsection{Temporal Distortions}
\label{sec:temporal-model}

To model temporal dynamics, we used a simplified variant of the \citet{horsager_predicting_2009} model, which incorporates two coupled leaky integrators to simulate neural desensitization $n(t)$ and phosphene brightness $b(t)$. The governing equations were:
\begin{align}
    \frac{dn(t)}{dt} &= -\tau_n n(t) + b_I(t), \label{eq:charge-accumulation}\\
    \frac{db(t)}{dt} &= -\tau_b b(t) - \alpha n(t) + b_I(t), \label{eq:phosphene-brightness}
\end{align}
where $b_I(t)$ was the instantaneous brightness (from the spatial model) calculated at time $t$.
Parameter values ($\tau_n = \SI{0.2}{\second}$, $\tau_b = \SI{5}{\second}$, and $\alpha = 0.2$) were fitted to reproduce temporal fading and persistence effects reported by Subject 5 of \citet{perez_fornos_temporal_2012} (see their Figure 4).

\subsubsection{Gaze-Contingent Phosphene Rendering}
\label{sec:gaze}

Modern retinal prostheses use head-mounted cameras, so the visual input remains stable in head-centered coordinates even as the eyes move. To simulate this realistically in VR, we implemented gaze-contingent rendering, ensuring that the simulated implant followed the participant's fixation point.

Using the HTC Vive Pro Eye, we tracked gaze in real time and shifted each video frame to center the implant on the current point of fixation. 
This rendered phosphenes in retinal coordinates, allowing stimulation patterns to move with the eye.
This is critical for capturing perceptual effects like fading, streaking, and local adaptation \cite{paraskevoudi_eye_2019,perez_fornos_temporal_2012}. 
Without this step, the simulation would unrealistically fix stimulation to the screen, producing smeared or distorted percepts during eye movements.

Gaze-contingent stimulation is increasingly seen as essential for biologically plausible SPV \citep{caspi_eye_2021,paraskevoudi_eye_2019,kasowski_simulated_2025}, and future prostheses are expected to support it either via onboard sensors or external eye trackers.

Our setup achieved mean eye-tracking precision of \SI{1.9}{\degree}, with \SI{94}{\percent} of samples falling within \SI{5}{\degree} of the target (Appendix~\ref{sec:app-eye-tracking}).

\begin{figure*}[t!]
    \centering
    \includegraphics[width=\linewidth]{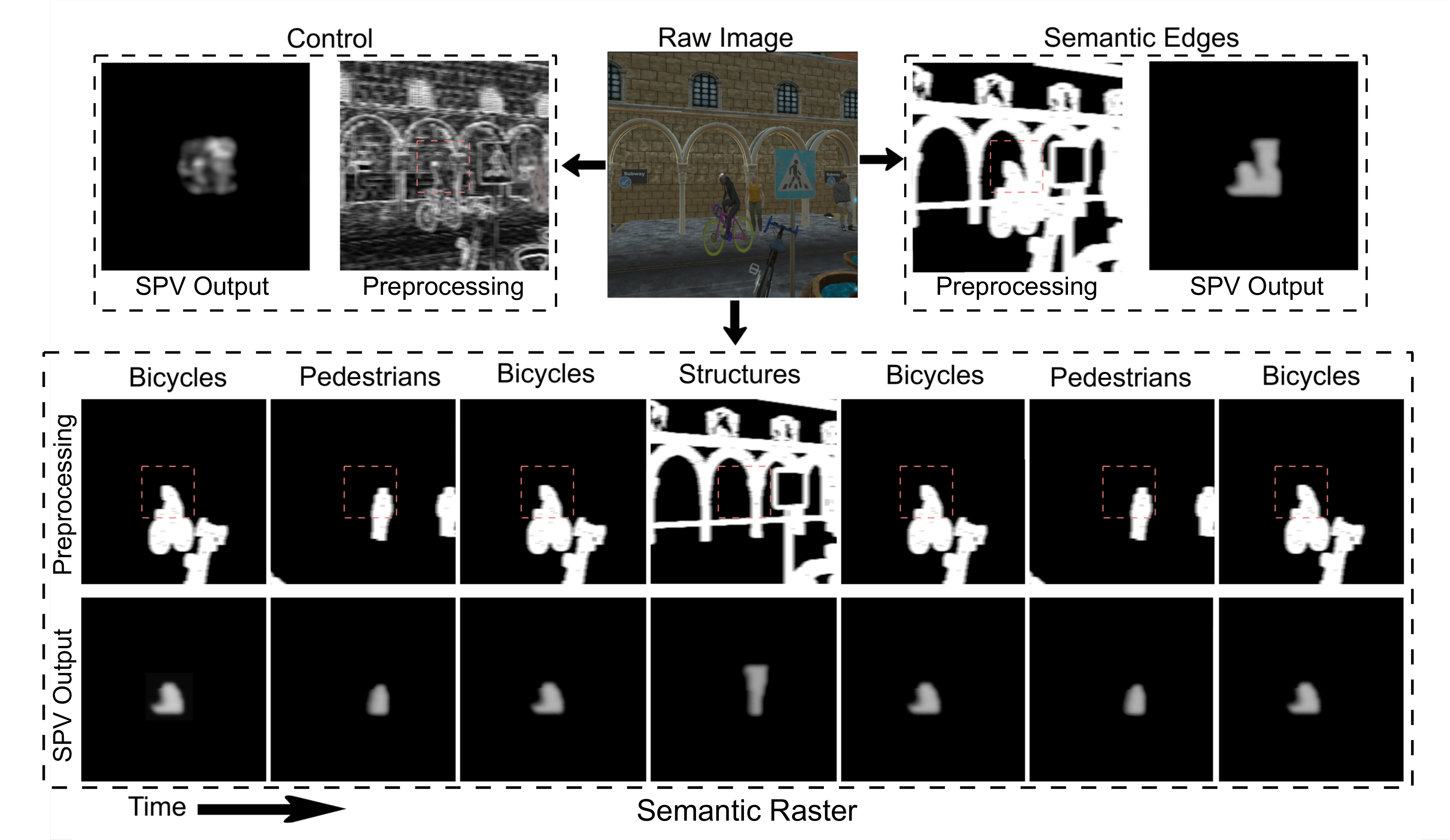}
    \caption{Scene simplification strategies tested in the study. The raw RGB image (top center) was processed using three methods. \emph{Control} (top left) applied a $3 \times 3$ Sobel filter to highlight edges without prioritizing task-relevant features. \emph{SemanticEdges} (top right) used semantic segmentation and a $7 \times 7$ kernel to enhance edges of selected classes (bicycles, pedestrians, and structures). \emph{SemanticRaster} (bottom) grouped semantic categories and displayed them sequentially over time, cycling semantic groups over time with higher update rates for dynamic objects. The final panel(s) for each method show(s) simulated prosthetic vision (SPV) output rendered in retinal coordinates; red dashed boxes mark the limited field of view of the SPV rendering.}
    \label{fig:methods-raster}
\end{figure*}

\subsection{Scene Simplification Strategies}

To evaluate the effect of different scene simplification strategies on wayfinding performance, the \ac{SPV} system rendered visual input using three distinct strategies:
\begin{enumerate}[topsep=0pt,itemsep=-1ex,partopsep=0pt,parsep=1ex,leftmargin=14pt,label=\roman*.]
    \item \emph{Control:} This baseline condition applied a standard $3 \times 3$ Sobel  kernel to the input for edge detection. While effective for emphasizing structural boundaries, this method lacked task-specific prioritization, often resulting in a cluttered visual field that could overwhelm users in complex environments.
    
    \item \emph{SemanticEdges:} A semantically informed edge filter that emphasized high-priority objects (e.g., pedestrians, bicycles, and structural features) based on scene understanding. A $7 \times 7$ Sobel kernel enhanced edges while suppressing irrelevant background details, reducing clutter and emphasizing salient features.

    \item \emph{SemanticRaster:} A novel strategy that combined semantic segmentation with temporal prioritization. Rather than displaying all object classes simultaneously, this mode cycled through key object categories over time (\SI{200}{\milli\second} per class), repeatedly displaying bicycles, then pedestrians, then structural edges. This schedule aimed to reduce crowding and improve perception under the low-resolution constraints of SPV (Fig.~\ref{fig:methods-raster}).
\end{enumerate}

\subsubsection{Task Relevance and Raster Schedule}
In our framework, an object class is considered \textit{task-relevant} if: (i) the task involves interacting with, avoiding, or locating that class, and (ii) failure to perceive the class negatively impacts performance (e.g., increased collision or timeout rates). 
These classes were identified with the help of a blind consultant and an orientation and mobility (O\&M) specialist.
For the present wayfinding task, this process yielded three key classes (i.e., bicycles, pedestrians, and structural edges) in that order of importance. 
The \emph{SemanticRaster} strategy reflected this priority by allocating equal temporal slots to each class (\SI{200}{\milli\second}), with more frequent recurrence of higher-ranked categories.

Importantly, this framework generalizes across tasks: A street-crossing scenario, for example, might prioritize cars and crosswalks, while an indoor task might emphasize doors and furniture. The rastering mechanism remains the same; only the set and ordering of semantic classes changes, based on structured input from end users and task pilots \cite{gamage_smart_2025,hoogsteen_beyond_2022}.

\subsubsection{Stimulation Constraints}
Although these strategies prioritized relevant information, they still required stimulating a large number of electrodes per frame---potentially exceeding safe limits. To mitigate this, all strategies employed a checkerboard raster pattern previously shown to be perceptually effective and safe \cite{kasowski_simulated_2025}. This pattern alternated activation across the electrode grid to prevent simultaneous stimulation of adjacent electrodes. Cycling at \SI{90}{\hertz} to match the headset's refresh rate, the approach exploited temporal integration to yield coherent percepts while minimizing crosstalk and phosphene fusion.

\subsection{Task \& Environment}

\begin{figure}[t!]
    \includegraphics[width=\linewidth]{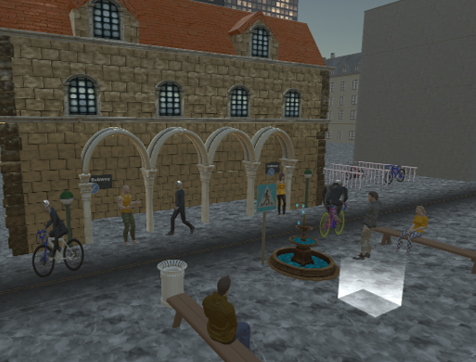}
    \caption{The simulated town square environment, highlighting the starting position in front of the fountain (white square). Participants were tasked with navigating to the left or right side of the subway station while avoiding collisions with pedestrians, bicycles, and static obstacles. Trials ended successfully when participants entered the correct side of the station or terminated early due to a collision with a biker or exceeding the \SI{50}{\second} time limit. }
    \label{fig:environment}
\end{figure}

Participants completed an ambulatory wayfinding task in a \ac{SPV} environment modeled after a $\SI{10}{\meter} \times \SI{10}{\meter}$ urban town square (Fig.~\ref{fig:environment}). 
The environment featured dynamic obstacles, such as bicycles and pedestrians, as well as static obstacles like benches and lampposts, accompanied by spatialized sound to replicate real-world navigation challenges.
The primary objective was to navigate from the starting position (in front of a central fountain) to one of two subway entrances (left or right) while avoiding collisions with obstacles.

Static obstacle configurations (e.g., benches, standing pedestrians) were drawn from a set of predefined layouts, with one pseudorandomly selected per trial to increase variability.
Dynamic obstacles followed predefined paths, but their speed and timing were randomized across trials to prevent memorization.
The SPV simulation reflected key constraints of current bionic vision systems, including a reduced visual field (\SI{14.6}{\degree} $\times$ \SI{14.6}{\degree}) and a phosphene resolution of $10 \times 10$ electrodes, rendered in a gaze-contingent and temporally dynamic manner.

To ensure participant safety during the ambulatory task, the virtual environment was overlaid onto a large, obstacle-free physical space. A trained experimenter continuously monitored participants and was ready to intervene if needed. 

\subsubsection{Training Phase}

\begin{figure}[t!]
    \includegraphics[width=\linewidth]{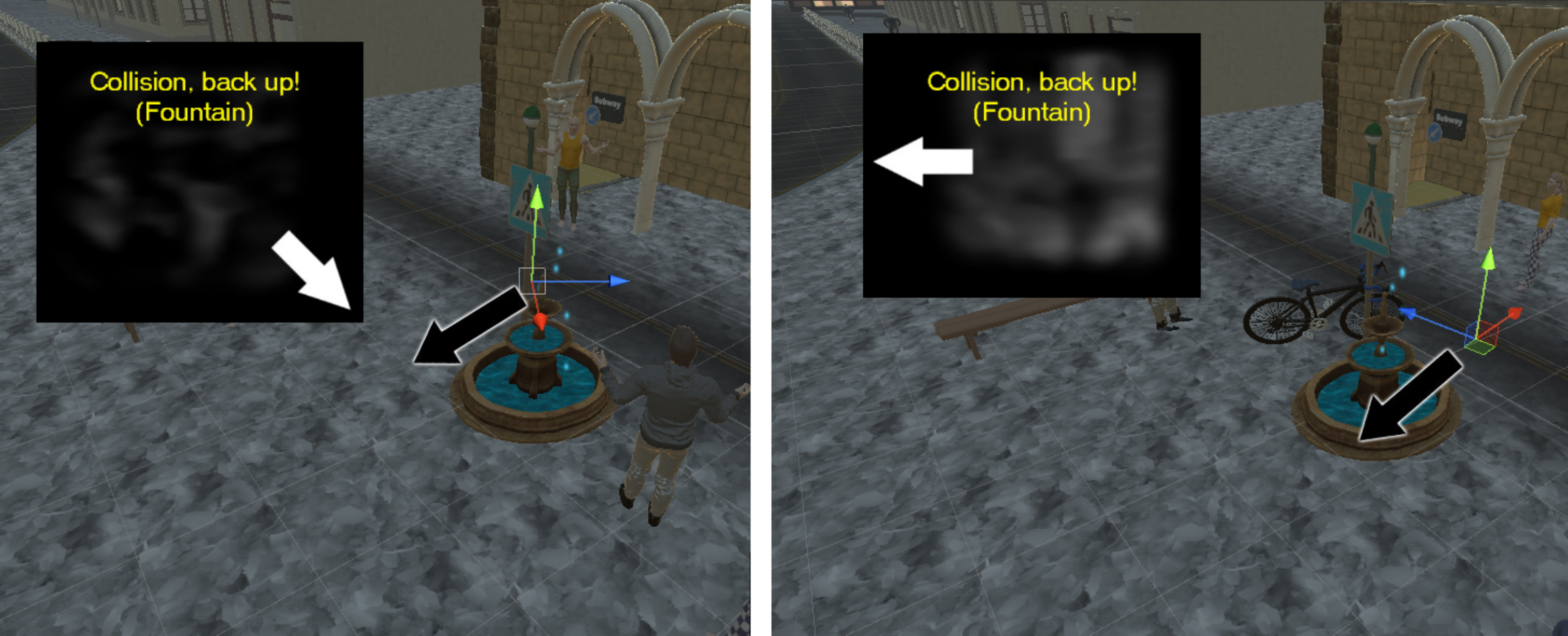}
    \caption{Collision feedback system. When a virtual collision was detected, participants were required to move \SI{0.25}{m} in the indicated direction to reset. A UI element displayed the message ``Collision, back up!'' along with the name of the collided object (e.g., ``Fountain'' or ``Person, walking''). An arrow indicated the required direction to move. Collisions could not be repeatedly triggered without moving back first, as a timeout mechanism prevented duplicate collision counts.}
    \label{fig:collisions}
\end{figure}

Participants completed a structured training session to acclimate to the SPV environment and task mechanics.
The session included five rounds in a simplified virtual scene with both static and dynamic obstacles.
The first four rounds used normal vision. The final round introduced SPV, including temporal distortions and one of the three scene simplification modes (Control, SemanticEdges, or SemanticRaster), corresponding to the participant’s upcoming block.
Participants practiced navigating and intentionally colliding with virtual objects (e.g., trashcans, bicycles) to experience the auditory and visual collision feedback.
To avoid double-counting, a cooldown period suppressed additional collision registration until the participant had moved at least \SI{0.25}{m} away (Fig.~\ref{fig:collisions}).

\subsubsection{Experimental Procedure}

The experiment followed a within-subjects block design, where participants completed all three  strategies (\emph{Control}, \emph{SemanticEdges}, and \emph{SemanticRaster}). 
The \emph{Control} condition was always presented first, while the order of the two smart strategies was counterbalanced across participants to control for learning effects. 
Each raster strategy constituted a block, with participants completing 10 trials per block, resulting in a total of 30 trials per participant.

Participants were tasked with the following objectives, in order of priority:
\begin{enumerate}[topsep=0pt,itemsep=-1ex,partopsep=0pt,parsep=1ex,leftmargin=14pt,label=\roman*.]
    \item reach the subway entrance within the time limit,
    \item avoid collisions with bicycles, and
    \item minimize other collisions.
\end{enumerate}

Each trial lasted up to 50 seconds and ended when the participant reached the target, collided with a bicycle (triggering a crashing sound), or ran out of time.
A countdown timer appeared with 10 seconds remaining to increase urgency.

\subsection{Data Collection \& Analysis}

Performance was assessed using the following metrics:
\begin{enumerate}[topsep=0pt,itemsep=-1ex,partopsep=0pt,parsep=1ex,leftmargin=14pt,label=\roman*.]
    \item \textbf{Task success:} Whether the participant reached the assigned subway entrance within the \SI{50}{\second} limit (bicycle collisions terminated the trial; other collisions did not).
    \item \textbf{Collision-free completion:} Indicator that the trial ended successfully \emph{and} registered zero collisions.
    \item \textbf{Collision rate:} Total number of collisions per trial, further broken down by obstacle type (stationary and moving).
    \item \textbf{Completion time:} Time to reach the target, computed for successful trials only.
    \item \textbf{Task difficulty:} Block-wise self-ratings on a 10-point Likert scale (1 = very easy, 10 = very hard).
\end{enumerate}

We analyzed the data using mixed-effects models to account for the repeated-measures design and individual variability. 
For each outcome measure, we included a fixed effect of scene simplification strategy (\texttt{Condition}: \emph{Control}, \emph{SemanticEdges}, \emph{SemanticRaster}) and a random intercept for each participant (\texttt{SubjectID}).
When applicable, we also included a centered trial index as a covariate to capture potential learning effects across the 30 trials. 
Interaction terms between \texttt{Condition} and trial index were retained only if they improved model fit, as assessed by likelihood ratio tests ($\alpha=.05$).

Binary outcomes (task success and collision-free completion) were analyzed using generalized linear mixed models (GLMMs) with a logit link:
\begin{equation*}
    \texttt{Success} \sim \texttt{Condition} + \texttt{TrialIndex} + (1 + \texttt{TrialIndex} \mid \texttt{SubjectID})
\end{equation*}

Count data (total collisions, stationary collisions, and moving collisions) were analyzed with Poisson GLMMs:
\begin{equation*}
    \texttt{Collisions} \sim \texttt{Condition} + \texttt{TrialIndex} + (1 \mid \texttt{SubjectID})    
\end{equation*}

Completion time for successful trials was analyzed using a linear mixed-effects model with Gaussian errors, including random slopes:
\begin{equation*}
    \texttt{Time} \sim \texttt{Condition} + \texttt{TrialIndex} + (1 + \texttt{TrialIndex} \mid \texttt{SubjectID})    
\end{equation*}

Difficulty ratings (on a 1--10 ordinal scale) were modeled using a cumulative link mixed model (logit link), including block presentation order as a fixed covariate:
\begin{equation*}
    \texttt{Difficulty} \sim \texttt{Condition} + \texttt{Order} + (1 \mid \texttt{SubjectID})    
\end{equation*}

All models were fit using R: \texttt{lme4} for linear and generalized linear models, and \texttt{ordinal::clmm} for cumulative link models. 
Post hoc contrasts were computed using the \texttt{emmeans} package, with Tukey adjustment for multiple comparisons.

\begin{figure*}[t!]
\includegraphics[width=\linewidth]{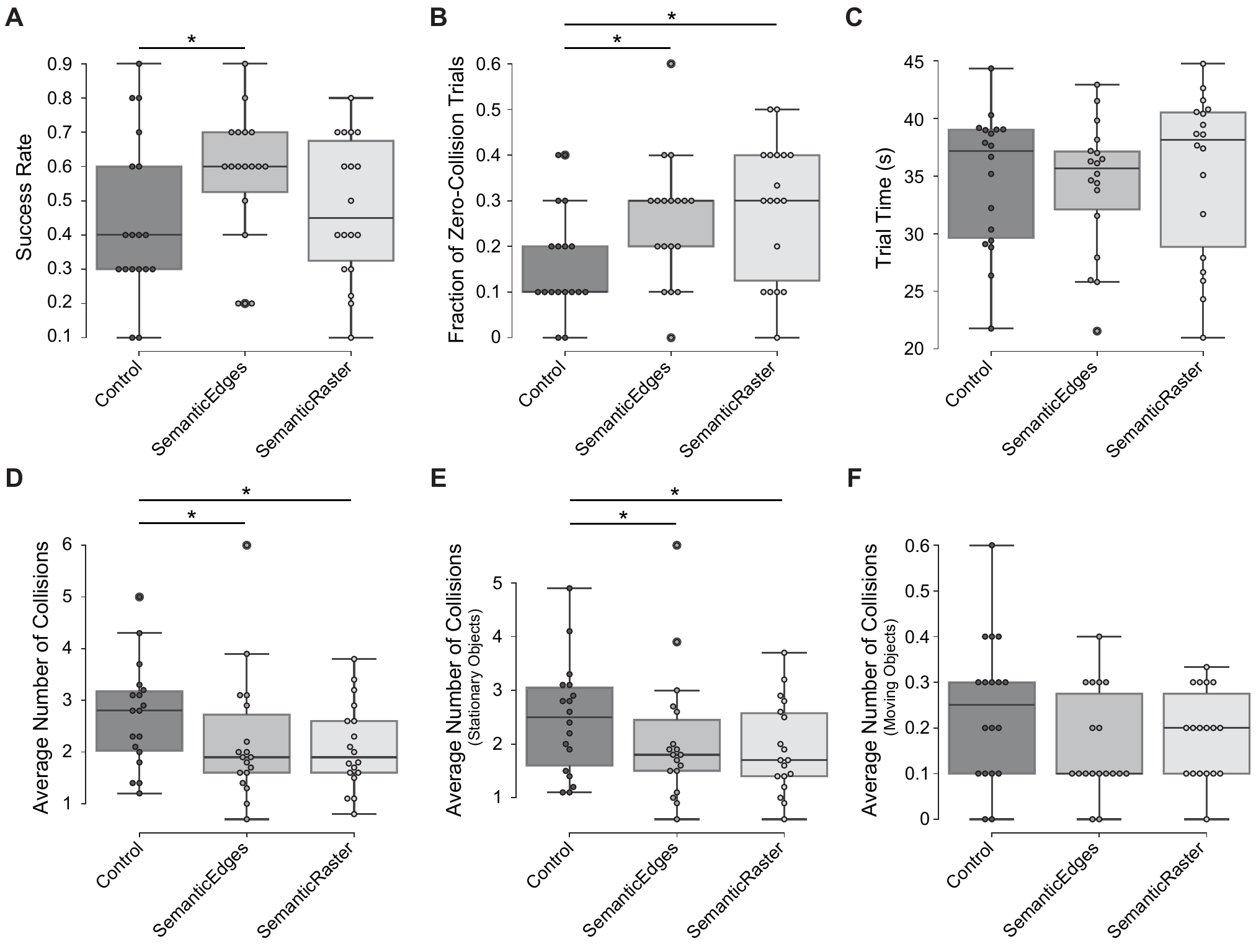}
\caption{Task performance metrics across conditions. 
    (A) Success rate: proportion of trials completed without collisions or time-outs. 
    (B) Fraction of successful zero-collision trials.
    (C) Trial completion time in seconds.
    (D) Average number of collisions per trial.
    (F) Average number of collisions involving static structures (e.g., fountain, bench, standing pedestrians).
    (E) Average number of collisions involving moving obstacles (i.e., bicycles).
    Each point represents a participant; boxplots show median, interquartile range, and range. Statistical significance between conditions is denoted by * ($p < .05$).}
\label{fig:results}
\end{figure*}

\section{Results}

\subsection{Task Success}

We first examined the impact of scene simplification on task success, defined as reaching the goal before the timer expired. A generalized linear mixed-effects model (GLMM) with fixed effects of \texttt{Condition} and centered \texttt{TrialIndex}, and by-subject random intercepts and learning slopes, revealed a significant benefit for the \emph{SemanticEdges} condition: participants were 1.84 times more likely to complete the task successfully compared to the \emph{Control} baseline ($\beta = 0.61 \pm 0.24$, $z = 2.59$, $p = .009$). 
\emph{SemanticRaster} showed a smaller, non-significant improvement ($\beta = 0.27$, $p = .24$).
There was a modest learning effect across trials ($\beta = 0.045$, $z = 3.99$, $p < .001$), but no significant interaction between condition and trial index ($\chi^2(2) = 0.17$, $p = .92$), suggesting that the scene simplification effects were stable over time.

To evaluate whether simplification also led to cleaner navigation, we examined the odds of completing a trial without any collisions. 
A separate GLMM (logit link) indicated that both \emph{SemanticEdges} and \emph{SemanticRaster} increased the likelihood of a clean run compared to the baseline (\emph{SemanticEdges}: OR = 1.8, $p = .086$; \emph{SemanticRaster}: OR = 2.1, $p = .018$), independent of trial index ($p > .9$). 
These results suggest that even when overall success rates are similar, \emph{SemanticRaster} may help users navigate more cleanly when successful.

Completion time did not vary significantly by condition or trial index (all $|t| < 1.3$, $p > .25$), indicating that these gains in accuracy were not simply due to participants slowing down.
Timeouts were rare, occurring on only 10 out of 540 trials ($<2\%$), further indicating that participants generally completed the task within the allotted time regardless of condition.

A \texttt{Condition} × \texttt{TrialIndex} interaction term was added to each model to test whether learning differed between strategies. Across all three outcomes (success, collision counts, trial time), the interaction was non-significant (all $p > .50$), indicating that participants improved (or plateaued) at comparable rates under \emph{SemanticRaster} and \emph{SemanticEdges}. 
Thus, \emph{SemanticRaster}'s cleaner-run advantage does not appear to hinge on an extended learning period.

\subsection{Collision Rates}

To better understand error patterns, we analyzed collision counts using Poisson GLMMs. 
Both \emph{SemanticEdges} and \emph{SemanticRaster} significantly reduced total collisions compared to \emph{Control}, by 21\% and 26\%, respectively (\emph{SemanticEdges}: $\beta = -0.236$, $p = .009$; \emph{SemanticRaster}: $\beta = -0.303$, $p = .001$). 
No significant difference emerged between the two smart strategies ($\beta = 0.067$, $p = .77$), and collision rates remained stable across trials ($p = .49$).

Breaking down collisions by object type revealed that these improvements were driven by reductions in contact with static obstacles. 
Both \emph{SemanticEdges} and \emph{SemanticRaster} significantly decreased stationary collisions relative to \emph{Control} (\emph{SemanticEdges}: –18\%, $\beta = -0.203$, $p = .035$; \emph{SemanticRaster}: –26\%, $\beta = -0.302$, $p = .003$). 
Again, the two smart modes did not differ significantly ($p = .61$), and there was no effect of trial index ($p = .12$).

Collisions with moving obstacles (e.g., cyclists) were rarer overall, but a marginal trend suggested that \emph{SemanticEdges} may reduce such collisions relative to baseline ($\beta = -0.443$, $p = .067$); the effect for \emph{SemanticRaster} was smaller and non-significant ($\beta = -0.324$, $p = .17$). 
Trial index showed a weak trend toward improvement ($p = .055$), but participant-level variance was negligible (singular fit). 
These results suggest that smart simplification is more effective for managing static than dynamic hazards.

Taken together, these findings indicate that improved performance was primarily driven by the simplification strategies themselves, rather than by learning across trials.

\subsection{Perceived Difficulty}

\begin{figure}[!ht]
    \centering
    \includegraphics[width=.8\linewidth]{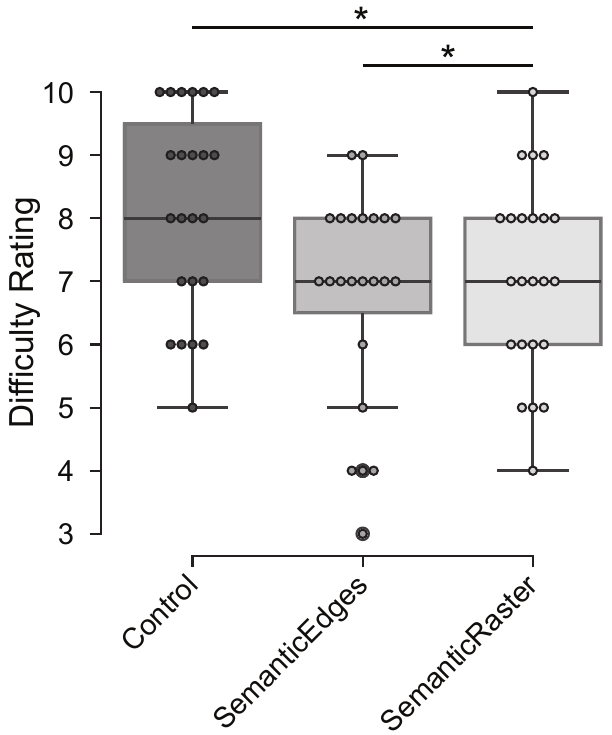}
    \caption{Difficulty ratings: Participants rated each condition's difficulty on a 10-point Likert scale (1 = very easy, 10 = very hard). Both smart strategies significantly reduced perceived difficulty compared to \emph{Control} (* \textbf{$p < .05$}).}
    \label{fig:results-difficulty}
\end{figure}

Finally, after completing all trials of a given condition, participants rated its difficulty on a 1--10 scale (Figure~\ref{fig:results-difficulty}). 
A cumulative link mixed model revealed a significant effect of condition: both \emph{SemanticEdges} ($\beta = -1.66$, $p = .004$) and \emph{SemanticRaster} ($\beta = -1.47$, $p = .010$) were perceived as less difficult than \emph{Control}. 
There was also a trend for later blocks to be rated as easier overall ($\beta = -0.48$, $p = .085$). 
Pairwise contrasts on the latent scale confirmed these findings: both smart modes were rated significantly easier than \emph{Control} (\emph{SemanticEdges}: $p = .012$; \emph{SemanticRaster}: $p = .027$), but did not differ from each other ($p = .94$). 
These subjective ratings align with the objective performance improvements observed above.

\section{Discussion}

This study evaluated two smart scene simplification strategies (\emph{SemanticEdges} and \emph{SemanticRaster}) designed to improve wayfinding performance in simulated prosthetic vision (SPV). 
Both strategies improved outcomes compared to a conventional pipeline (\emph{Control}), but in distinct ways: \emph{SemanticEdges} increased the odds of success, while \emph{SemanticRaster} boosted the likelihood of collision-free completions. 
These findings underscore the value of semantically driven preprocessing, while highlighting the trade-offs and complementary benefits of different display strategies.

\subsection{Complementary Roles of Static and Temporal Simplification}

Our data suggest that a static overlay of all task‑relevant object classes provides strong global awareness (i.e., participants finished more trials successfully, Fig.~\ref{fig:results}A) whereas sequencing those classes over time trades some global context for reduced clutter, resulting in fewer collisions (Fig.~\ref{fig:results}B) and lower self‑reported effort (Fig.~\ref{fig:results-difficulty}).
This makes intuitive sense: presenting everything at once maximizes information density but risks visual crowding, while staggered presentation lightens instantaneous load but can momentarily hide context.  
Which trade‑off is preferable appears to depend on what ``failure'' looks like in the task: missing an exit (\emph{SemanticEdges} helps) vs. clipping a hazard (\emph{SemanticRaster} helps).

Crucially, neither strategy harmed performance relative to baseline, and participants acclimated within a handful of trials. Across thirty trials, we observed only modest learning effects, and subjective difficulty ratings dropped by 1.5--2 points relative to the baseline for both smart strategies (Fig.~\ref{fig:results-difficulty}).
This rapid uptake is encouraging because perceptual learning in actual implant users is often measured in weeks or months \cite{stronks_functional_2014,dagnelie_performance_2016,erickson-davis_what_2021}. 


\subsection{Relevance for Temporally Multiplexed Implants}

Modern retinal prostheses already rely on raster‑like stimulation because safety limits restrict how many electrodes may fire simultaneously \citep{luo_argusr_2016}.  
Our \emph{SemanticRaster} shows that the same temporal budget required for safe charge delivery can be repurposed to convey task semantics: instead of sweeping the retinal array in a fixed spatial pattern, firmware could sweep \textit{through} semantic layers (e.g., hazards $\rightarrow$ landmarks $\rightarrow$ context).  
Although our SPV experiment used a simple three‑layer schedule, the underlying idea (content‑aware time‑division) aligns with recent calls for ``adaptive stimulation policies'' that are tailored to task and context \citep{beyeler_towards_2022}.

To be clear, we are not suggesting that the three object categories used here are universally optimal. They were chosen for this specific navigation task and urban environment, guided by input from a blind consultant and an O\&M specialist. In practice, both the set of semantic categories and their ordering should be tailored to the user's needs and the situational context. 
A co-design study with end users would be essential to determine appropriate categories, priorities, and update rates for different applications.

\subsection{Connections to Bandwidth‑Limited XR}

Similar constraints arise in head‑worn AR, remote telepresence, and low‑vision aids: pixel budgets are often constrained not just by hardware limitations (e.g., microdisplay resolution, battery life, or wireless throughput) but also by the limits of human attention \citep{patney_perceptually-based_2016,guenter_foveated_2012}. 
To cope, XR systems typically reduce resolution in the periphery, drop frames, or stream sparse features such as keypoints or depth edges when networks degrade or scenes become complex.
Our results add empirical evidence that temporally multiplexing entire semantic layers, rather than just lowering spatial resolution, can be viable when clutter is the bottleneck.
This temporal allocation of bandwidth may be especially useful in low-vision aids, where even modest increases in scene complexity can overwhelm the user \citep{lieby_substituting_2011,holiel_pre-processing_nodate,kasowski_systematic_2023}.

\subsection{Why Simulate Bionic Vision in Sighted Participants?}

Given that no FDA-approved or commercially available bionic eye exists today, there is currently no large user base for systematic study. Clinical studies typically involve only a handful of implanted users due to medical, logistical, and financial constraints. As a result, simulated prosthetic vision (SPV) has emerged as a widely accepted and cost-effective method for evaluating encoding strategies in controlled experiments \cite{boyle_region--interest_2008,li_image_2018,mccarthy_mobility_2014}.

Our approach builds on this foundation but moves beyond traditional SPV methods, which often rely on simplistic, idealized phosphenes (e.g., round Gaussian blobs) rendered frame-by-frame \cite{dagnelie_real_2007,chen_simulating_2009,lui_transformative_2012,sanchez-garcia_indoor_2019}. 
%
While sighted participants cannot capture the long-term neural adaptation experienced by implant users, they remain a valuable population for early-stage, within-subject comparisons \cite{beyeler_learning_2017}. 
Virtual prototyping can help identify promising strategies and avoid costly design missteps before real-world clinical deployment \cite{beyeler_towards_2022}.

\subsection{Limitations \& Future Directions}

While this study advances the understanding of semantic scene simplification strategies, several limitations must be acknowledged that should guide future work.

First, we examined only one specific navigation task in a controlled virtual setting. Further work is needed to evaluate how these findings generalize across tasks, environments, and user needs. 
Second, while our simulation incorporates key perceptual features like fading and distortion, it does not account for the full range of individual variability observed in real prosthesis users, including variability in electrode-retina interactions or cortical plasticity \cite{matteucci_effect_2016,legge_low_2016}. 
Finally, dynamic obstacles remain a challenge: none of the strategies tested here significantly reduced collisions with moving objects, and future work should explore motion-aware or adaptive encoding techniques to address this gap.

Despite these limitations, our results support the value of semantic simplification (and especially temporal structuring) as a means to reduce perceptual burden in prosthetic vision. 
Continued progress towards developing intelligent, adaptive bionic vision systems~\cite{beyeler_towards_2022,lozano_neurolight_2020,rao_brain_2020} will require closing the loop between simulation and clinical deployment, ideally through collaborative studies with implanted users \cite{nadolskis_aligning_2024,erickson-davis_what_2021}. 
Combining advanced computer vision techniques with adaptive, multimodal feedback systems holds the promise of empowering users to navigate complex environments with greater independence and confidence.

\section{Conclusion}

We compared two semantic preprocessing strategies for prosthetic vision: one that highlights all task-relevant objects simultaneously (\emph{SemanticEdges}) and another that sequences them over time (\emph{SemanticRaster}). Both approaches improved wayfinding and user experience over a traditional edge-based baseline, with \emph{SemanticEdges} aiding global awareness and \emph{SemanticRaster} reducing collisions in dynamic scenes. These results underscore the value of temporally adaptive, task-informed encoding for visual prostheses and suggest design principles for clutter-aware XR interfaces.



\bibliographystyle{ACM-Reference-Format}
\bibliography{references}


\begin{thebibliography}{64}


\ifx \showCODEN    \undefined \def \showCODEN     #1{\unskip}     \fi
\ifx \showDOI      \undefined \def \showDOI       #1{#1}\fi
\ifx \showISBNx    \undefined \def \showISBNx     #1{\unskip}     \fi
\ifx \showISBNxiii \undefined \def \showISBNxiii  #1{\unskip}     \fi
\ifx \showISSN     \undefined \def \showISSN      #1{\unskip}     \fi
\ifx \showLCCN     \undefined \def \showLCCN      #1{\unskip}     \fi
\ifx \shownote     \undefined \def \shownote      #1{#1}          \fi
\ifx \showarticletitle \undefined \def \showarticletitle #1{#1}   \fi
\ifx \showURL      \undefined \def \showURL       {\relax}        \fi
\providecommand\bibfield[2]{#2}
\providecommand\bibinfo[2]{#2}
\providecommand\natexlab[1]{#1}
\providecommand\showeprint[2][]{arXiv:#2}

\bibitem[Beyeler et~al\mbox{.}(2019)]%
        {beyeler_model_2019}
\bibfield{author}{\bibinfo{person}{Michael Beyeler}, \bibinfo{person}{Devyani Nanduri}, \bibinfo{person}{James~D. Weiland}, \bibinfo{person}{Ariel Rokem}, \bibinfo{person}{Geoffrey~M. Boynton}, {and} \bibinfo{person}{Ione Fine}.} \bibinfo{year}{2019}\natexlab{}.
\newblock \showarticletitle{A model of ganglion axon pathways accounts for percepts elicited by retinal implants}.
\newblock \bibinfo{journal}{\emph{Scientific Reports}} \bibinfo{volume}{9}, \bibinfo{number}{1} (\bibinfo{date}{June} \bibinfo{year}{2019}), \bibinfo{pages}{1--16}.
\newblock
\showISSN{2045-2322}
\urldef\tempurl%
\url{https://doi.org/10.1038/s41598-019-45416-4}
\showDOI{\tempurl}


\bibitem[Beyeler et~al\mbox{.}(2017)]%
        {beyeler_learning_2017}
\bibfield{author}{\bibinfo{person}{M. Beyeler}, \bibinfo{person}{A. Rokem}, \bibinfo{person}{G.~M. Boynton}, {and} \bibinfo{person}{I. Fine}.} \bibinfo{year}{2017}\natexlab{}.
\newblock \showarticletitle{Learning to see again: biological constraints on cortical plasticity and the implications for sight restoration technologies}.
\newblock \bibinfo{journal}{\emph{J Neural Eng}} \bibinfo{volume}{14}, \bibinfo{number}{5} (\bibinfo{date}{June} \bibinfo{year}{2017}), \bibinfo{pages}{051003}.
\newblock
\showISSN{1741-2552 (Electronic) 1741-2552 (Linking)}
\urldef\tempurl%
\url{https://doi.org/10.1088/1741-2552/aa795e}
\showDOI{\tempurl}


\bibitem[Beyeler and Sanchez-Garcia(2022)]%
        {beyeler_towards_2022}
\bibfield{author}{\bibinfo{person}{Michael Beyeler} {and} \bibinfo{person}{Melani Sanchez-Garcia}.} \bibinfo{year}{2022}\natexlab{}.
\newblock \showarticletitle{Towards a {Smart} {Bionic} {Eye}: {AI}-powered artificial vision for the treatment of incurable blindness}.
\newblock \bibinfo{journal}{\emph{Journal of Neural Engineering}} \bibinfo{volume}{19}, \bibinfo{number}{6} (\bibinfo{date}{Dec.} \bibinfo{year}{2022}), \bibinfo{pages}{063001}.
\newblock
\showISSN{1741-2552}
\urldef\tempurl%
\url{https://doi.org/10.1088/1741-2552/aca69d}
\showDOI{\tempurl}
\newblock
\shownote{Publisher: IOP Publishing}.


\bibitem[Bourne et~al\mbox{.}(2020)]%
        {bourne_global_2020}
\bibfield{author}{\bibinfo{person}{Rupert R.~A. Bourne}, \bibinfo{person}{Jaimie Adelson}, \bibinfo{person}{Seth Flaxman}, \bibinfo{person}{Paul Briant}, \bibinfo{person}{Michele Bottone}, \bibinfo{person}{Theo Vos}, \bibinfo{person}{Kovin Naidoo}, \bibinfo{person}{Tasanee Braithwaite}, \bibinfo{person}{Maria Cicinelli}, \bibinfo{person}{Jost Jonas}, \bibinfo{person}{Hans Limburg}, \bibinfo{person}{Serge Resnikoff}, \bibinfo{person}{Alex Silvester}, \bibinfo{person}{Vinay Nangia}, {and} \bibinfo{person}{Hugh~R. Taylor}.} \bibinfo{year}{2020}\natexlab{}.
\newblock \showarticletitle{Global {Prevalence} of {Blindness} and {Distance} and {Near} {Vision} {Impairment} in 2020: progress towards the {Vision} 2020 targets and what the future holds.}
\newblock \bibinfo{journal}{\emph{Investigative Ophthalmology \& Visual Science}} \bibinfo{volume}{61}, \bibinfo{number}{7} (\bibinfo{date}{June} \bibinfo{year}{2020}), \bibinfo{pages}{2317--2317}.
\newblock
\showISSN{1552-5783}
\urldef\tempurl%
\url{https://iovs.arvojournals.org/article.aspx?articleid=2767477}
\showURL{%
\tempurl}


\bibitem[Boyle et~al\mbox{.}(2008)]%
        {boyle_region--interest_2008}
\bibfield{author}{\bibinfo{person}{Justin~R. Boyle}, \bibinfo{person}{Anthony~J. Maeder}, {and} \bibinfo{person}{Wageeh~W. Boles}.} \bibinfo{year}{2008}\natexlab{}.
\newblock \showarticletitle{Region-of-interest processing for electronic visual prostheses}.
\newblock \bibinfo{journal}{\emph{Journal of Electronic Imaging}} \bibinfo{volume}{17}, \bibinfo{number}{1} (\bibinfo{date}{Jan.} \bibinfo{year}{2008}), \bibinfo{pages}{013002}.
\newblock
\showISSN{1017-9909, 1560-229X}
\urldef\tempurl%
\url{https://doi.org/10.1117/1.2841708}
\showDOI{\tempurl}
\newblock
\shownote{Publisher: International Society for Optics and Photonics}.


\bibitem[Caspi et~al\mbox{.}(2021)]%
        {caspi_eye_2021}
\bibfield{author}{\bibinfo{person}{Avi Caspi}, \bibinfo{person}{Michael~P. Barry}, \bibinfo{person}{Uday~K. Patel}, \bibinfo{person}{Michelle~Armenta Salas}, \bibinfo{person}{Jessy~D. Dorn}, \bibinfo{person}{Arup Roy}, \bibinfo{person}{Soroush Niketeghad}, \bibinfo{person}{Robert~J. Greenberg}, {and} \bibinfo{person}{Nader Pouratian}.} \bibinfo{year}{2021}\natexlab{}.
\newblock \showarticletitle{Eye movements and the perceived location of phosphenes generated by intracranial primary visual cortex stimulation in the blind}.
\newblock \bibinfo{journal}{\emph{Brain Stimulation}} \bibinfo{volume}{14}, \bibinfo{number}{4} (\bibinfo{date}{July} \bibinfo{year}{2021}), \bibinfo{pages}{851--860}.
\newblock
\showISSN{1935-861X}
\urldef\tempurl%
\url{https://doi.org/10.1016/j.brs.2021.04.019}
\showDOI{\tempurl}


\bibitem[Chen et~al\mbox{.}(2009)]%
        {chen_simulating_2009}
\bibfield{author}{\bibinfo{person}{S.~C. Chen}, \bibinfo{person}{G.~J. Suaning}, \bibinfo{person}{J.~W. Morley}, {and} \bibinfo{person}{N.~H. Lovell}.} \bibinfo{year}{2009}\natexlab{}.
\newblock \showarticletitle{Simulating prosthetic vision: {I}. {Visual} models of phosphenes}.
\newblock \bibinfo{journal}{\emph{Vision Research}} \bibinfo{volume}{49}, \bibinfo{number}{12} (\bibinfo{date}{June} \bibinfo{year}{2009}), \bibinfo{pages}{1493--506}.
\newblock
\showISSN{1878-5646 (Electronic) 0042-6989 (Linking)}


\bibitem[Chen et~al\mbox{.}(2020)]%
        {chen_shape_2020}
\bibfield{author}{\bibinfo{person}{Xing Chen}, \bibinfo{person}{Feng Wang}, \bibinfo{person}{Eduardo Fernandez}, {and} \bibinfo{person}{Pieter~R. Roelfsema}.} \bibinfo{year}{2020}\natexlab{}.
\newblock \showarticletitle{Shape perception via a high-channel-count neuroprosthesis in monkey visual cortex}.
\newblock \bibinfo{journal}{\emph{Science}} \bibinfo{volume}{370}, \bibinfo{number}{6521} (\bibinfo{date}{Dec.} \bibinfo{year}{2020}), \bibinfo{pages}{1191--1196}.
\newblock
\showISSN{0036-8075, 1095-9203}
\urldef\tempurl%
\url{https://doi.org/10.1126/science.abd7435}
\showDOI{\tempurl}
\newblock
\shownote{Publisher: American Association for the Advancement of Science Section: Research Article}.


\bibitem[Chung et~al\mbox{.}(2024)]%
        {chung_large-scale_2024}
\bibfield{author}{\bibinfo{person}{Daeun~Joyce Chung}, \bibinfo{person}{Muya Guoji}, \bibinfo{person}{Nina Mindel}, \bibinfo{person}{Alexis Malkin}, \bibinfo{person}{Fernando Alberotrio}, \bibinfo{person}{Shane Lowe}, \bibinfo{person}{Chris McNally}, \bibinfo{person}{Casandra Xavier}, {and} \bibinfo{person}{Paul Ruvolo}.} \bibinfo{year}{2024}\natexlab{}.
\newblock \bibinfo{title}{Large-scale, {Longitudinal}, {Hybrid} {Participatory} {Design} {Program} to {Create} {Navigation} {Technology} for the {Blind}}.
\newblock
\newblock
\urldef\tempurl%
\url{https://doi.org/10.48550/arXiv.2410.00192}
\showDOI{\tempurl}
\newblock
\shownote{arXiv:2410.00192}.


\bibitem[Dagnelie et~al\mbox{.}(2016)]%
        {dagnelie_performance_2016}
\bibfield{author}{\bibinfo{person}{G. Dagnelie}, \bibinfo{person}{P. Christopher}, \bibinfo{person}{A. Arditi}, \bibinfo{person}{L. da Cruz}, \bibinfo{person}{J.~L. Duncan}, \bibinfo{person}{A.~C. Ho}, \bibinfo{person}{L.~C. de Koo}, \bibinfo{person}{J.~A. Sahel}, \bibinfo{person}{P.~E. Stanga}, \bibinfo{person}{G. Thumann}, \bibinfo{person}{Y. Wang}, \bibinfo{person}{M. Arsiero}, \bibinfo{person}{J.~D. Dorn}, \bibinfo{person}{R.~J. Greenberg}, {and} \bibinfo{person}{I.~I. Study~Group Argus}.} \bibinfo{year}{2016}\natexlab{}.
\newblock \showarticletitle{Performance of real-world functional vision tasks by blind subjects improves after implantation with the {Argus}({R}) {II} retinal prosthesis system}.
\newblock \bibinfo{journal}{\emph{Clin Experiment Ophthalmol}} (\bibinfo{date}{Aug.} \bibinfo{year}{2016}).
\newblock
\showISSN{1442-9071 (Electronic) 1442-6404 (Linking)}
\urldef\tempurl%
\url{https://doi.org/10.1111/ceo.12812}
\showDOI{\tempurl}


\bibitem[Dagnelie et~al\mbox{.}(2007)]%
        {dagnelie_real_2007}
\bibfield{author}{\bibinfo{person}{G. Dagnelie}, \bibinfo{person}{P. Keane}, \bibinfo{person}{V. Narla}, \bibinfo{person}{L. Yang}, \bibinfo{person}{J. Weiland}, {and} \bibinfo{person}{M. Humayun}.} \bibinfo{year}{2007}\natexlab{}.
\newblock \showarticletitle{Real and virtual mobility performance in simulated prosthetic vision}.
\newblock \bibinfo{journal}{\emph{J Neural Eng}} \bibinfo{volume}{4}, \bibinfo{number}{1} (\bibinfo{date}{March} \bibinfo{year}{2007}), \bibinfo{pages}{S92--101}.
\newblock
\showISSN{1741-2560 (Print) 1741-2552 (Linking)}
\urldef\tempurl%
\url{https://doi.org/10.1088/1741-2560/4/1/S11}
\showDOI{\tempurl}


\bibitem[Erickson-Davis and Korzybska(2021)]%
        {erickson-davis_what_2021}
\bibfield{author}{\bibinfo{person}{Cordelia Erickson-Davis} {and} \bibinfo{person}{Helma Korzybska}.} \bibinfo{year}{2021}\natexlab{}.
\newblock \showarticletitle{What do blind people “see” with retinal prostheses? {Observations} and qualitative reports of epiretinal implant users}.
\newblock \bibinfo{journal}{\emph{PLOS ONE}} \bibinfo{volume}{16}, \bibinfo{number}{2} (\bibinfo{date}{Feb.} \bibinfo{year}{2021}), \bibinfo{pages}{e0229189}.
\newblock
\showISSN{1932-6203}
\urldef\tempurl%
\url{https://doi.org/10.1371/journal.pone.0229189}
\showDOI{\tempurl}
\newblock
\shownote{Publisher: Public Library of Science}.


\bibitem[Fernandez(2018)]%
        {fernandez_development_2018}
\bibfield{author}{\bibinfo{person}{Eduardo Fernandez}.} \bibinfo{year}{2018}\natexlab{}.
\newblock \showarticletitle{Development of visual {Neuroprostheses}: trends and challenges}.
\newblock \bibinfo{journal}{\emph{Bioelectronic Medicine}} \bibinfo{volume}{4}, \bibinfo{number}{1} (\bibinfo{date}{Aug.} \bibinfo{year}{2018}), \bibinfo{pages}{12}.
\newblock
\showISSN{2332-8886}
\urldef\tempurl%
\url{https://doi.org/10.1186/s42234-018-0013-8}
\showDOI{\tempurl}


\bibitem[Fernández et~al\mbox{.}(2021)]%
        {fernandez_visual_2021}
\bibfield{author}{\bibinfo{person}{Eduardo Fernández}, \bibinfo{person}{Arantxa Alfaro}, \bibinfo{person}{Cristina Soto-Sánchez}, \bibinfo{person}{Pablo Gonzalez-Lopez}, \bibinfo{person}{Antonio~M. Lozano}, \bibinfo{person}{Sebastian Peña}, \bibinfo{person}{Maria~Dolores Grima}, \bibinfo{person}{Alfonso Rodil}, \bibinfo{person}{Bernardeta Gómez}, \bibinfo{person}{Xing Chen}, \bibinfo{person}{Pieter~R. Roelfsema}, \bibinfo{person}{John~D. Rolston}, \bibinfo{person}{Tyler~S. Davis}, {and} \bibinfo{person}{Richard~A. Normann}.} \bibinfo{year}{2021}\natexlab{}.
\newblock \showarticletitle{Visual percepts evoked with an intracortical 96-channel microelectrode array inserted in human occipital cortex}.
\newblock \bibinfo{journal}{\emph{The Journal of Clinical Investigation}} \bibinfo{volume}{131}, \bibinfo{number}{23} (\bibinfo{date}{Dec.} \bibinfo{year}{2021}), \bibinfo{pages}{e151331}.
\newblock
\showISSN{1558-8238}
\urldef\tempurl%
\url{https://doi.org/10.1172/JCI151331}
\showDOI{\tempurl}


\bibitem[Fine and Boynton(2024)]%
        {fine_virtual_2024}
\bibfield{author}{\bibinfo{person}{Ione Fine} {and} \bibinfo{person}{Geoffrey~M. Boynton}.} \bibinfo{year}{2024}\natexlab{}.
\newblock \showarticletitle{A virtual patient simulation modeling the neural and perceptual effects of human visual cortical stimulation, from pulse trains to percepts}.
\newblock \bibinfo{journal}{\emph{Scientific Reports}} \bibinfo{volume}{14}, \bibinfo{number}{1} (\bibinfo{date}{July} \bibinfo{year}{2024}), \bibinfo{pages}{17400}.
\newblock
\showISSN{2045-2322}
\urldef\tempurl%
\url{https://doi.org/10.1038/s41598-024-65337-1}
\showDOI{\tempurl}
\newblock
\shownote{Publisher: Nature Publishing Group}.


\bibitem[Gamage et~al\mbox{.}(2025)]%
        {gamage_smart_2025}
\bibfield{author}{\bibinfo{person}{Bhanuka Gamage}, \bibinfo{person}{Nicola McDowell}, \bibinfo{person}{Dijana Kovacic}, \bibinfo{person}{Leona Holloway}, \bibinfo{person}{Thanh-Toan Do}, \bibinfo{person}{Nicholas Price}, \bibinfo{person}{Arthur Lowery}, {and} \bibinfo{person}{Kim Marriott}.} \bibinfo{year}{2025}\natexlab{}.
\newblock \bibinfo{title}{Smart {Glasses} for {CVI}: {Co}-{Designing} {Extended} {Reality} {Solutions} to {Support} {Environmental} {Perception} by {People} with {Cerebral} {Visual} {Impairment}}.
\newblock
\newblock
\urldef\tempurl%
\url{https://doi.org/10.48550/arXiv.2506.19210}
\showDOI{\tempurl}
\newblock
\shownote{arXiv:2506.19210 [cs]}.


\bibitem[Geruschat et~al\mbox{.}(2016)]%
        {geruschat_analysis_2016}
\bibfield{author}{\bibinfo{person}{Duane~R Geruschat}, \bibinfo{person}{Thomas~P Richards}, \bibinfo{person}{Aries Arditi}, \bibinfo{person}{Lyndon da Cruz}, \bibinfo{person}{Gislin Dagnelie}, \bibinfo{person}{Jessy~D Dorn}, \bibinfo{person}{Jacque~L Duncan}, \bibinfo{person}{Allen~C Ho}, \bibinfo{person}{Lisa~C Olmos~de Koo}, \bibinfo{person}{José‐Alain Sahel}, \bibinfo{person}{Paulo~E Stanga}, \bibinfo{person}{Gabriele Thumann}, \bibinfo{person}{Vizhong Wang}, {and} \bibinfo{person}{Robert~J Greenberg}.} \bibinfo{year}{2016}\natexlab{}.
\newblock \showarticletitle{An analysis of observer‐rated functional vision in patients implanted with the {Argus} {II} {Retinal} {Prosthesis} {System} at three years}.
\newblock \bibinfo{journal}{\emph{Clinical \& Experimental Optometry}} \bibinfo{volume}{99}, \bibinfo{number}{3} (\bibinfo{date}{May} \bibinfo{year}{2016}), \bibinfo{pages}{227--232}.
\newblock
\showISSN{0816-4622}
\urldef\tempurl%
\url{https://doi.org/10.1111/cxo.12359}
\showDOI{\tempurl}


\bibitem[Granley and Beyeler(2021)]%
        {granley_computational_2021}
\bibfield{author}{\bibinfo{person}{Jacob Granley} {and} \bibinfo{person}{Michael Beyeler}.} \bibinfo{year}{2021}\natexlab{}.
\newblock \showarticletitle{A {Computational} {Model} of {Phosphene} {Appearance} for {Epiretinal} {Prostheses}}. In \bibinfo{booktitle}{\emph{2021 43rd {Annual} {International} {Conference} of the {IEEE} {Engineering} in {Medicine} \& {Biology} {Society} ({EMBC})}}. \bibinfo{pages}{4477--4481}.
\newblock
\urldef\tempurl%
\url{https://doi.org/10.1109/EMBC46164.2021.9629663}
\showDOI{\tempurl}
\newblock
\shownote{ISSN: 2694-0604}.


\bibitem[Guenter et~al\mbox{.}(2012)]%
        {guenter_foveated_2012}
\bibfield{author}{\bibinfo{person}{Brian Guenter}, \bibinfo{person}{Mark Finch}, \bibinfo{person}{Steven Drucker}, \bibinfo{person}{Desney Tan}, {and} \bibinfo{person}{John Snyder}.} \bibinfo{year}{2012}\natexlab{}.
\newblock \showarticletitle{Foveated {3D} graphics}.
\newblock \bibinfo{journal}{\emph{ACM Trans. Graph.}} \bibinfo{volume}{31}, \bibinfo{number}{6} (\bibinfo{date}{Nov.} \bibinfo{year}{2012}), \bibinfo{pages}{164:1--164:10}.
\newblock
\showISSN{0730-0301}
\urldef\tempurl%
\url{https://doi.org/10.1145/2366145.2366183}
\showDOI{\tempurl}


\bibitem[Han et~al\mbox{.}(2021)]%
        {han_deep_2021}
\bibfield{author}{\bibinfo{person}{Nicole Han}, \bibinfo{person}{Sudhanshu Srivastava}, \bibinfo{person}{Aiwen Xu}, \bibinfo{person}{Devi Klein}, {and} \bibinfo{person}{Michael Beyeler}.} \bibinfo{year}{2021}\natexlab{}.
\newblock \showarticletitle{Deep {Learning}–{Based} {Scene} {Simplification} for {Bionic} {Vision}}. In \bibinfo{booktitle}{\emph{Augmented {Humans} {Conference} 2021}} \emph{(\bibinfo{series}{{AHs}'21})}. \bibinfo{publisher}{Association for Computing Machinery}, \bibinfo{address}{New York, NY, USA}, \bibinfo{pages}{45--54}.
\newblock
\showISBNx{978-1-4503-8428-5}
\urldef\tempurl%
\url{https://doi.org/10.1145/3458709.3458982}
\showDOI{\tempurl}


\bibitem[Hayes et~al\mbox{.}(2003)]%
        {hayes_visually_2003}
\bibfield{author}{\bibinfo{person}{J.~S. Hayes}, \bibinfo{person}{V.~T. Yin}, \bibinfo{person}{D. Piyathaisere}, \bibinfo{person}{J.~D. Weiland}, \bibinfo{person}{M.~S. Humayun}, {and} \bibinfo{person}{G. Dagnelie}.} \bibinfo{year}{2003}\natexlab{}.
\newblock \showarticletitle{Visually guided performance of simple tasks using simulated prosthetic vision}.
\newblock \bibinfo{journal}{\emph{Artif Organs}} \bibinfo{volume}{27}, \bibinfo{number}{11} (\bibinfo{date}{Nov.} \bibinfo{year}{2003}), \bibinfo{pages}{1016--28}.
\newblock
\showISSN{0160-564X (Print) 0160-564X (Linking)}


\bibitem[Holiel et~al\mbox{.}({[n.\,d.]})]%
        {holiel_pre-processing_nodate}
\bibfield{author}{\bibinfo{person}{Heidi~Ahmed Holiel}, \bibinfo{person}{Sahar~Ali Fawzi}, {and} \bibinfo{person}{Walid Al-Atabany}.} \bibinfo{year}{[n.\,d.]}\natexlab{}.
\newblock \showarticletitle{Pre-processing visual scenes for retinal prosthesis systems: {A} comprehensive review}.
\newblock \bibinfo{journal}{\emph{Artificial Organs}} \bibinfo{volume}{n/a}, \bibinfo{number}{n/a} (\bibinfo{year}{[n.\,d.]}).
\newblock
\showISSN{1525-1594}
\urldef\tempurl%
\url{https://doi.org/10.1111/aor.14824}
\showDOI{\tempurl}
\newblock
\shownote{\_eprint: https://onlinelibrary.wiley.com/doi/pdf/10.1111/aor.14824}.


\bibitem[Hoogsteen et~al\mbox{.}(2022)]%
        {hoogsteen_beyond_2022}
\bibfield{author}{\bibinfo{person}{Karst~M.P. Hoogsteen}, \bibinfo{person}{Sarit Szpiro}, \bibinfo{person}{Gabriel Kreiman}, {and} \bibinfo{person}{Eli Peli}.} \bibinfo{year}{2022}\natexlab{}.
\newblock \showarticletitle{Beyond the {Cane}: {Describing} {Urban} {Scenes} to {Blind} {People} for {Mobility} {Tasks}}.
\newblock \bibinfo{journal}{\emph{ACM Transactions on Accessible Computing}} (\bibinfo{date}{Feb.} \bibinfo{year}{2022}).
\newblock
\showISSN{1936-7228}
\urldef\tempurl%
\url{https://doi.org/10.1145/3522757}
\showDOI{\tempurl}
\newblock
\shownote{Just Accepted}.


\bibitem[Horsager et~al\mbox{.}(2009)]%
        {horsager_predicting_2009}
\bibfield{author}{\bibinfo{person}{A. Horsager}, \bibinfo{person}{S.~H. Greenwald}, \bibinfo{person}{J.~D. Weiland}, \bibinfo{person}{M.~S. Humayun}, \bibinfo{person}{R.~J. Greenberg}, \bibinfo{person}{M.~J. McMahon}, \bibinfo{person}{G.~M. Boynton}, {and} \bibinfo{person}{I. Fine}.} \bibinfo{year}{2009}\natexlab{}.
\newblock \showarticletitle{Predicting visual sensitivity in retinal prosthesis patients}.
\newblock \bibinfo{journal}{\emph{Invest Ophthalmol Vis Sci}} \bibinfo{volume}{50}, \bibinfo{number}{4} (\bibinfo{date}{April} \bibinfo{year}{2009}), \bibinfo{pages}{1483--91}.
\newblock
\showISSN{1552-5783 (Electronic) 0146-0404 (Linking)}
\urldef\tempurl%
\url{https://doi.org/10.1167/iovs.08-2595}
\showDOI{\tempurl}


\bibitem[Hou et~al\mbox{.}(2024a)]%
        {hou_axonal_2024}
\bibfield{author}{\bibinfo{person}{Yuchen Hou}, \bibinfo{person}{Devyani Nanduri}, \bibinfo{person}{Jacob Granley}, \bibinfo{person}{James~D. Weiland}, {and} \bibinfo{person}{Michael Beyeler}.} \bibinfo{year}{2024}\natexlab{a}.
\newblock \showarticletitle{Axonal stimulation affects the linear summation of single-point perception in three {Argus} {II} users}.
\newblock \bibinfo{journal}{\emph{Journal of Neural Engineering}} \bibinfo{volume}{21}, \bibinfo{number}{2} (\bibinfo{date}{April} \bibinfo{year}{2024}), \bibinfo{pages}{026031}.
\newblock
\showISSN{1741-2552}
\urldef\tempurl%
\url{https://doi.org/10.1088/1741-2552/ad31c4}
\showDOI{\tempurl}
\newblock
\shownote{Publisher: IOP Publishing}.


\bibitem[Hou et~al\mbox{.}(2024b)]%
        {hou_predicting_2024}
\bibfield{author}{\bibinfo{person}{Yuchen Hou}, \bibinfo{person}{Laya Pullela}, \bibinfo{person}{Jiaxin Su}, \bibinfo{person}{Sriya Aluru}, \bibinfo{person}{Shivani Sista}, \bibinfo{person}{Xiankun Lu}, {and} \bibinfo{person}{Michael Beyeler}.} \bibinfo{year}{2024}\natexlab{b}.
\newblock \showarticletitle{Predicting the {Temporal} {Dynamics} of {Prosthetic} {Vision}}. In \bibinfo{booktitle}{\emph{2024 46th {Annual} {International} {Conference} of the {IEEE} {Engineering} in {Medicine} and {Biology} {Society} ({EMBC})}}. \bibinfo{pages}{1--4}.
\newblock
\urldef\tempurl%
\url{https://doi.org/10.1109/EMBC53108.2024.10782668}
\showDOI{\tempurl}
\newblock
\shownote{ISSN: 2694-0604}.


\bibitem[Jung et~al\mbox{.}(2024)]%
        {jung_stable_2024}
\bibfield{author}{\bibinfo{person}{Taesung Jung}, \bibinfo{person}{Nanyu Zeng}, \bibinfo{person}{Jason~D. Fabbri}, \bibinfo{person}{Guy Eichler}, \bibinfo{person}{Zhe Li}, \bibinfo{person}{Konstantin Willeke}, \bibinfo{person}{Katie~E. Wingel}, \bibinfo{person}{Agrita Dubey}, \bibinfo{person}{Rizwan Huq}, \bibinfo{person}{Mohit Sharma}, \bibinfo{person}{Yaoxing Hu}, \bibinfo{person}{Girish Ramakrishnan}, \bibinfo{person}{Kevin Tien}, \bibinfo{person}{Paolo Mantovani}, \bibinfo{person}{Abhinav Parihar}, \bibinfo{person}{Heyu Yin}, \bibinfo{person}{Denise Oswalt}, \bibinfo{person}{Alexander Misdorp}, \bibinfo{person}{Ilke Uguz}, \bibinfo{person}{Tori Shinn}, \bibinfo{person}{Gabrielle~J. Rodriguez}, \bibinfo{person}{Cate Nealley}, \bibinfo{person}{Ian Gonzales}, \bibinfo{person}{Michael Roukes}, \bibinfo{person}{Jeffrey Knecht}, \bibinfo{person}{Daniel Yoshor}, \bibinfo{person}{Peter Canoll}, \bibinfo{person}{Eleonora Spinazzi}, \bibinfo{person}{Luca~P. Carloni}, \bibinfo{person}{Bijan Pesaran},
  \bibinfo{person}{Saumil Patel}, \bibinfo{person}{Brett Youngerman}, \bibinfo{person}{R.~James Cotton}, \bibinfo{person}{Andreas Tolias}, {and} \bibinfo{person}{Kenneth~L. Shepard}.} \bibinfo{year}{2024}\natexlab{}.
\newblock \bibinfo{title}{Stable, chronic in-vivo recordings from a fully wireless subdural-contained 65,536-electrode brain-computer interface device}.
\newblock
\newblock
\urldef\tempurl%
\url{https://doi.org/10.1101/2024.05.17.594333}
\showDOI{\tempurl}
\newblock
\shownote{Pages: 2024.05.17.594333 Section: New Results}.


\bibitem[Kasowski and Beyeler(2022)]%
        {kasowski_immersive_2022}
\bibfield{author}{\bibinfo{person}{Justin Kasowski} {and} \bibinfo{person}{Michael Beyeler}.} \bibinfo{year}{2022}\natexlab{}.
\newblock \showarticletitle{Immersive {Virtual} {Reality} {Simulations} of {Bionic} {Vision}}. In \bibinfo{booktitle}{\emph{Augmented {Humans} 2022}} \emph{(\bibinfo{series}{{AHs} 2022})}. \bibinfo{publisher}{Association for Computing Machinery}, \bibinfo{address}{New York, NY, USA}, \bibinfo{pages}{82--93}.
\newblock
\showISBNx{978-1-4503-9632-5}
\urldef\tempurl%
\url{https://doi.org/10.1145/3519391.3522752}
\showDOI{\tempurl}


\bibitem[Kasowski et~al\mbox{.}(2023)]%
        {kasowski_systematic_2023}
\bibfield{author}{\bibinfo{person}{Justin Kasowski}, \bibinfo{person}{Byron~A. Johnson}, \bibinfo{person}{Ryan Neydavood}, \bibinfo{person}{Anvitha Akkaraju}, {and} \bibinfo{person}{Michael Beyeler}.} \bibinfo{year}{2023}\natexlab{}.
\newblock \showarticletitle{A systematic review of extended reality ({XR}) for understanding and augmenting vision loss}.
\newblock \bibinfo{journal}{\emph{Journal of Vision}} \bibinfo{volume}{23}, \bibinfo{number}{5} (\bibinfo{date}{May} \bibinfo{year}{2023}), \bibinfo{pages}{5}.
\newblock
\showISSN{1534-7362}
\urldef\tempurl%
\url{https://doi.org/10.1167/jov.23.5.5}
\showDOI{\tempurl}


\bibitem[Kasowski et~al\mbox{.}(2025)]%
        {kasowski_simulated_2025}
\bibfield{author}{\bibinfo{person}{Justin~M Kasowski}, \bibinfo{person}{Apurv Varshney}, \bibinfo{person}{Roksana Sadeghi}, {and} \bibinfo{person}{Michael Beyeler}.} \bibinfo{year}{2025}\natexlab{}.
\newblock \showarticletitle{Simulated prosthetic vision confirms checkerboard as an effective raster pattern for epiretinal implants}.
\newblock \bibinfo{journal}{\emph{Journal of Neural Engineering}} (\bibinfo{year}{2025}).
\newblock
\showISSN{1741-2552}
\urldef\tempurl%
\url{https://doi.org/10.1088/1741-2552/adecc4}
\showDOI{\tempurl}


\bibitem[Legge and Chung(2016)]%
        {legge_low_2016}
\bibfield{author}{\bibinfo{person}{Gordon~E. Legge} {and} \bibinfo{person}{Susana~T.L. Chung}.} \bibinfo{year}{2016}\natexlab{}.
\newblock \showarticletitle{Low {Vision} and {Plasticity}: {Implications} for {Rehabilitation}}.
\newblock \bibinfo{journal}{\emph{Annual Review of Vision Science}} \bibinfo{volume}{2}, \bibinfo{number}{1} (\bibinfo{date}{Oct.} \bibinfo{year}{2016}), \bibinfo{pages}{321--343}.
\newblock
\showISSN{2374-4642, 2374-4650}
\urldef\tempurl%
\url{https://doi.org/10.1146/annurev-vision-111815-114344}
\showDOI{\tempurl}


\bibitem[Li et~al\mbox{.}(2018)]%
        {li_image_2018}
\bibfield{author}{\bibinfo{person}{Heng Li}, \bibinfo{person}{Xiaofan Su}, \bibinfo{person}{Jing Wang}, \bibinfo{person}{Han Kan}, \bibinfo{person}{Tingting Han}, \bibinfo{person}{Yajie Zeng}, {and} \bibinfo{person}{Xinyu Chai}.} \bibinfo{year}{2018}\natexlab{}.
\newblock \showarticletitle{Image processing strategies based on saliency segmentation for object recognition under simulated prosthetic vision}.
\newblock \bibinfo{journal}{\emph{Artificial Intelligence in Medicine}}  \bibinfo{volume}{84} (\bibinfo{date}{Jan.} \bibinfo{year}{2018}), \bibinfo{pages}{64--78}.
\newblock
\showISSN{0933-3657}
\urldef\tempurl%
\url{https://doi.org/10.1016/j.artmed.2017.11.001}
\showDOI{\tempurl}


\bibitem[Lieby et~al\mbox{.}(2011)]%
        {lieby_substituting_2011}
\bibfield{author}{\bibinfo{person}{Paulette Lieby}, \bibinfo{person}{Nick Barnes}, \bibinfo{person}{Chris McCarthy}, \bibinfo{person}{Nianjun Liu}, \bibinfo{person}{Hugh Dennett}, \bibinfo{person}{Janine~G. Walker}, \bibinfo{person}{Viorica Botea}, {and} \bibinfo{person}{Adele~F. Scott}.} \bibinfo{year}{2011}\natexlab{}.
\newblock \showarticletitle{Substituting depth for intensity and real-time phosphene rendering: visual navigation under low vision conditions}. In \bibinfo{booktitle}{\emph{Annual {International} {Conference} of the {IEEE} {Engineering} in {Medicine} and {Biology} {Society}. {IEEE} {Engineering} in {Medicine} and {Biology} {Society}. {Annual} {International} {Conference}}}, Vol.~\bibinfo{volume}{2011}. \bibinfo{pages}{8017--8020}.
\newblock
\urldef\tempurl%
\url{https://doi.org/10.1109/IEMBS.2011.6091977}
\showDOI{\tempurl}


\bibitem[Lorach et~al\mbox{.}(2015)]%
        {lorach_photovoltaic_2015}
\bibfield{author}{\bibinfo{person}{H. Lorach}, \bibinfo{person}{G. Goetz}, \bibinfo{person}{R. Smith}, \bibinfo{person}{X. Lei}, \bibinfo{person}{Y. Mandel}, \bibinfo{person}{T. Kamins}, \bibinfo{person}{K. Mathieson}, \bibinfo{person}{P. Huie}, \bibinfo{person}{J. Harris}, \bibinfo{person}{A. Sher}, {and} \bibinfo{person}{D. Palanker}.} \bibinfo{year}{2015}\natexlab{}.
\newblock \showarticletitle{Photovoltaic restoration of sight with high visual acuity}.
\newblock \bibinfo{journal}{\emph{Nat Med}} \bibinfo{volume}{21}, \bibinfo{number}{5} (\bibinfo{date}{May} \bibinfo{year}{2015}), \bibinfo{pages}{476--82}.
\newblock
\showISSN{1546-170X (Electronic) 1078-8956 (Linking)}
\urldef\tempurl%
\url{https://doi.org/10.1038/nm.3851}
\showDOI{\tempurl}


\bibitem[Lozano et~al\mbox{.}(2020)]%
        {lozano_neurolight_2020}
\bibfield{author}{\bibinfo{person}{Antonio Lozano}, \bibinfo{person}{Juan~Sebastian Suarez}, \bibinfo{person}{Cristina Soto-Sanchez}, \bibinfo{person}{Javier Garrigos}, \bibinfo{person}{J.~Javier Martinez-Alvarez}, \bibinfo{person}{J.~Manuel Ferrandez}, {and} \bibinfo{person}{Eduardo Fernandez}.} \bibinfo{year}{2020}\natexlab{}.
\newblock \showarticletitle{Neurolight: {A} {Deep} {Learning} {Neural} {Interface} for {Cortical} {Visual} {Prostheses}}.
\newblock \bibinfo{journal}{\emph{International Journal of Neural Systems}} (\bibinfo{date}{May} \bibinfo{year}{2020}).
\newblock
\showISSN{0129-0657}
\urldef\tempurl%
\url{https://doi.org/10.1142/S0129065720500458}
\showDOI{\tempurl}
\newblock
\shownote{Publisher: World Scientific Publishing Co.}.


\bibitem[Lui et~al\mbox{.}(2012)]%
        {lui_transformative_2012}
\bibfield{author}{\bibinfo{person}{Wen Lik~Dennis Lui}, \bibinfo{person}{Damien Browne}, \bibinfo{person}{Lindsay Kleeman}, \bibinfo{person}{Tom Drummond}, {and} \bibinfo{person}{Wai~Ho Li}.} \bibinfo{year}{2012}\natexlab{}.
\newblock \showarticletitle{Transformative {Reality}: improving bionic vision with robotic sensing}. In \bibinfo{booktitle}{\emph{Annual {International} {Conference} of the {IEEE} {Engineering} in {Medicine} and {Biology} {Society}. {IEEE} {Engineering} in {Medicine} and {Biology} {Society}. {Annual} {International} {Conference}}}, Vol.~\bibinfo{volume}{2012}. \bibinfo{pages}{304--307}.
\newblock
\urldef\tempurl%
\url{https://doi.org/10.1109/EMBC.2012.6345929}
\showDOI{\tempurl}


\bibitem[Luo and da~Cruz(2016)]%
        {luo_argusr_2016}
\bibfield{author}{\bibinfo{person}{Y.~H. Luo} {and} \bibinfo{person}{L. da Cruz}.} \bibinfo{year}{2016}\natexlab{}.
\newblock \showarticletitle{The {Argus}(({R})) {II} {Retinal} {Prosthesis} {System}}.
\newblock \bibinfo{journal}{\emph{Prog Retin Eye Res}}  \bibinfo{volume}{50} (\bibinfo{date}{Jan.} \bibinfo{year}{2016}), \bibinfo{pages}{89--107}.
\newblock
\showISSN{1873-1635 (Electronic) 1350-9462 (Linking)}
\urldef\tempurl%
\url{https://doi.org/10.1016/j.preteyeres.2015.09.003}
\showDOI{\tempurl}


\bibitem[Matteucci et~al\mbox{.}(2016)]%
        {matteucci_effect_2016}
\bibfield{author}{\bibinfo{person}{Paul~B. Matteucci}, \bibinfo{person}{Alejandro Barriga-Rivera}, \bibinfo{person}{Calvin~D. Eiber}, \bibinfo{person}{Nigel~H. Lovell}, \bibinfo{person}{John~W. Morley}, {and} \bibinfo{person}{Gregg~J. Suaning}.} \bibinfo{year}{2016}\natexlab{}.
\newblock \showarticletitle{The {Effect} of {Electric} {Cross}-{Talk} in {Retinal} {Neurostimulation}}.
\newblock \bibinfo{journal}{\emph{Investigative Ophthalmology \& Visual Science}} \bibinfo{volume}{57}, \bibinfo{number}{3} (\bibinfo{date}{March} \bibinfo{year}{2016}), \bibinfo{pages}{1031--1037}.
\newblock
\showISSN{1552-5783}
\urldef\tempurl%
\url{https://doi.org/10.1167/iovs.15-18400}
\showDOI{\tempurl}


\bibitem[McCarthy et~al\mbox{.}(2014)]%
        {mccarthy_mobility_2014}
\bibfield{author}{\bibinfo{person}{Chris McCarthy}, \bibinfo{person}{Janine~G. Walker}, \bibinfo{person}{Paulette Lieby}, \bibinfo{person}{Adele Scott}, {and} \bibinfo{person}{Nick Barnes}.} \bibinfo{year}{2014}\natexlab{}.
\newblock \showarticletitle{Mobility and low contrast trip hazard avoidance using augmented depth}.
\newblock \bibinfo{journal}{\emph{Journal of Neural Engineering}} \bibinfo{volume}{12}, \bibinfo{number}{1} (\bibinfo{date}{Nov.} \bibinfo{year}{2014}), \bibinfo{pages}{016003}.
\newblock
\showISSN{1741-2552}
\urldef\tempurl%
\url{https://doi.org/10.1088/1741-2560/12/1/016003}
\showDOI{\tempurl}
\newblock
\shownote{Publisher: IOP Publishing}.


\bibitem[Musk and Neuralink(2019)]%
        {musk_integrated_2019}
\bibfield{author}{\bibinfo{person}{Elon Musk} {and} \bibinfo{person}{Neuralink}.} \bibinfo{year}{2019}\natexlab{}.
\newblock \showarticletitle{An {Integrated} {Brain}-{Machine} {Interface} {Platform} {With} {Thousands} of {Channels}}.
\newblock \bibinfo{journal}{\emph{Journal of Medical Internet Research}} \bibinfo{volume}{21}, \bibinfo{number}{10} (\bibinfo{date}{Oct.} \bibinfo{year}{2019}), \bibinfo{pages}{e16194}.
\newblock
\urldef\tempurl%
\url{https://doi.org/10.2196/16194}
\showDOI{\tempurl}
\newblock
\shownote{Company: Journal of Medical Internet Research Distributor: Journal of Medical Internet Research Institution: Journal of Medical Internet Research Label: Journal of Medical Internet Research Publisher: JMIR Publications Inc., Toronto, Canada}.


\bibitem[Nadolskis et~al\mbox{.}(2024)]%
        {nadolskis_aligning_2024}
\bibfield{author}{\bibinfo{person}{Lucas Nadolskis}, \bibinfo{person}{Lily~M. Turkstra}, \bibinfo{person}{Ebenezer Larnyo}, {and} \bibinfo{person}{Michael Beyeler}.} \bibinfo{year}{2024}\natexlab{}.
\newblock \showarticletitle{Aligning {Visual} {Prosthetic} {Development} {With} {Implantee} {Needs}}.
\newblock \bibinfo{journal}{\emph{Translational Vision Science \& Technology}} \bibinfo{volume}{13}, \bibinfo{number}{11} (\bibinfo{date}{Nov.} \bibinfo{year}{2024}), \bibinfo{pages}{28}.
\newblock
\showISSN{2164-2591}
\urldef\tempurl%
\url{https://doi.org/10.1167/tvst.13.11.28}
\showDOI{\tempurl}


\bibitem[Palanker et~al\mbox{.}(2020)]%
        {palanker_photovoltaic_2020}
\bibfield{author}{\bibinfo{person}{Daniel Palanker}, \bibinfo{person}{Yannick Le~Mer}, \bibinfo{person}{Saddek Mohand-Said}, \bibinfo{person}{Mahiul Muqit}, {and} \bibinfo{person}{Jose~A. Sahel}.} \bibinfo{year}{2020}\natexlab{}.
\newblock \showarticletitle{Photovoltaic {Restoration} of {Central} {Vision} in {Atrophic} {Age}-{Related} {Macular} {Degeneration}}.
\newblock \bibinfo{journal}{\emph{Ophthalmology}} (\bibinfo{date}{Feb.} \bibinfo{year}{2020}).
\newblock
\showISSN{1549-4713}
\urldef\tempurl%
\url{https://doi.org/10.1016/j.ophtha.2020.02.024}
\showDOI{\tempurl}


\bibitem[Paraskevoudi and Pezaris(2019)]%
        {paraskevoudi_eye_2019}
\bibfield{author}{\bibinfo{person}{Nadia Paraskevoudi} {and} \bibinfo{person}{John~S. Pezaris}.} \bibinfo{year}{2019}\natexlab{}.
\newblock \showarticletitle{Eye {Movement} {Compensation} and {Spatial} {Updating} in {Visual} {Prosthetics}: {Mechanisms}, {Limitations} and {Future} {Directions}}.
\newblock \bibinfo{journal}{\emph{Frontiers in Systems Neuroscience}}  \bibinfo{volume}{12} (\bibinfo{year}{2019}).
\newblock
\showISSN{1662-5137}
\urldef\tempurl%
\url{https://doi.org/10.3389/fnsys.2018.00073}
\showDOI{\tempurl}


\bibitem[Patney et~al\mbox{.}(2016)]%
        {patney_perceptually-based_2016}
\bibfield{author}{\bibinfo{person}{Anjul Patney}, \bibinfo{person}{Joohwan Kim}, \bibinfo{person}{Marco Salvi}, \bibinfo{person}{Anton Kaplanyan}, \bibinfo{person}{Chris Wyman}, \bibinfo{person}{Nir Benty}, \bibinfo{person}{Aaron Lefohn}, {and} \bibinfo{person}{David Luebke}.} \bibinfo{year}{2016}\natexlab{}.
\newblock \showarticletitle{Perceptually-based foveated virtual reality}. In \bibinfo{booktitle}{\emph{{ACM} {SIGGRAPH} 2016 {Emerging} {Technologies}}} \emph{(\bibinfo{series}{{SIGGRAPH} '16})}. \bibinfo{publisher}{Association for Computing Machinery}, \bibinfo{address}{New York, NY, USA}, \bibinfo{pages}{1--2}.
\newblock
\showISBNx{978-1-4503-4372-5}
\urldef\tempurl%
\url{https://doi.org/10.1145/2929464.2929472}
\showDOI{\tempurl}


\bibitem[Perez-Yus et~al\mbox{.}(2017)]%
        {perez-yus_depth_2017}
\bibfield{author}{\bibinfo{person}{Alejandro Perez-Yus}, \bibinfo{person}{Jesus Bermudez-Cameo}, \bibinfo{person}{Gonzalo Lopez-Nicolas}, {and} \bibinfo{person}{Jose~J. Guerrero}.} \bibinfo{year}{2017}\natexlab{}.
\newblock \showarticletitle{Depth and {Motion} {Cues} {With} {Phosphene} {Patterns} for {Prosthetic} {Vision}}. \bibinfo{pages}{1516--1525}.
\newblock
\urldef\tempurl%
\url{http://openaccess.thecvf.com/content_ICCV_2017_workshops/w22/html/Perez-Yus_Depth_and_Motion_ICCV_2017_paper.html}
\showURL{%
\tempurl}


\bibitem[Pérez~Fornos et~al\mbox{.}(2012)]%
        {perez_fornos_temporal_2012}
\bibfield{author}{\bibinfo{person}{Angélica Pérez~Fornos}, \bibinfo{person}{Jörg Sommerhalder}, \bibinfo{person}{Lyndon da Cruz}, \bibinfo{person}{Jose~Alain Sahel}, \bibinfo{person}{Saddek Mohand-Said}, \bibinfo{person}{Farhad Hafezi}, {and} \bibinfo{person}{Marco Pelizzone}.} \bibinfo{year}{2012}\natexlab{}.
\newblock \showarticletitle{Temporal {Properties} of {Visual} {Perception} on {Electrical} {Stimulation} of the {Retina}}.
\newblock \bibinfo{journal}{\emph{Investigative Ophthalmology \& Visual Science}} \bibinfo{volume}{53}, \bibinfo{number}{6} (\bibinfo{year}{2012}), \bibinfo{pages}{2720--2731}.
\newblock
\showISSN{1552-5783}
\urldef\tempurl%
\url{https://doi.org/10.1167/iovs.11-9344}
\showDOI{\tempurl}


\bibitem[Rao(2020)]%
        {rao_brain_2020}
\bibfield{author}{\bibinfo{person}{Rajesh P.~N. Rao}.} \bibinfo{year}{2020}\natexlab{}.
\newblock \showarticletitle{Brain {Co}-{Processors}: {Using} {AI} to {Restore} and {Augment} {Brain} {Function}}.
\newblock  (\bibinfo{date}{Dec.} \bibinfo{year}{2020}).
\newblock
\urldef\tempurl%
\url{https://arxiv.org/abs/2012.03378v1}
\showURL{%
\tempurl}


\bibitem[Rasla and Beyeler(2022)]%
        {rasla_relative_2022}
\bibfield{author}{\bibinfo{person}{Alex Rasla} {and} \bibinfo{person}{Michael Beyeler}.} \bibinfo{year}{2022}\natexlab{}.
\newblock \showarticletitle{The {Relative} {Importance} of {Depth} {Cues} and {Semantic} {Edges} for {Indoor} {Mobility} {Using} {Simulated} {Prosthetic} {Vision} in {Immersive} {Virtual} {Reality}}. In \bibinfo{booktitle}{\emph{Proceedings of the 28th {ACM} {Symposium} on {Virtual} {Reality} {Software} and {Technology}}} \emph{(\bibinfo{series}{{VRST} '22})}. \bibinfo{publisher}{Association for Computing Machinery}, \bibinfo{address}{New York, NY, USA}, \bibinfo{pages}{1--11}.
\newblock
\showISBNx{978-1-4503-9889-3}
\urldef\tempurl%
\url{https://doi.org/10.1145/3562939.3565620}
\showDOI{\tempurl}


\bibitem[Reis et~al\mbox{.}(2011)]%
        {reis_patient_2011}
\bibfield{author}{\bibinfo{person}{Catarina~I. Reis}, \bibinfo{person}{Carla~S. Freire}, \bibinfo{person}{Joaquin Fernández}, {and} \bibinfo{person}{Josep~M. Monguet}.} \bibinfo{year}{2011}\natexlab{}.
\newblock \showarticletitle{Patient {Centered} {Design}: {Challenges} and {Lessons} {Learned} from {Working} with {Health} {Professionals} and {Schizophrenic} {Patients} in e-{Therapy} {Contexts}}. In \bibinfo{booktitle}{\emph{{ENTERprise} {Information} {Systems}}} \emph{(\bibinfo{series}{Communications in {Computer} and {Information} {Science}})}, \bibfield{editor}{\bibinfo{person}{Maria~Manuela Cruz-Cunha}, \bibinfo{person}{João Varajão}, \bibinfo{person}{Philip Powell}, {and} \bibinfo{person}{Ricardo Martinho}} (Eds.). \bibinfo{publisher}{Springer}, \bibinfo{address}{Berlin, Heidelberg}, \bibinfo{pages}{1--10}.
\newblock
\showISBNx{978-3-642-24352-3}
\urldef\tempurl%
\url{https://doi.org/10.1007/978-3-642-24352-3_1}
\showDOI{\tempurl}


\bibitem[Rizzo et~al\mbox{.}(2003)]%
        {rizzo_perceptual_2003}
\bibfield{author}{\bibinfo{person}{J.~F. Rizzo}, \bibinfo{person}{J. Wyatt}, \bibinfo{person}{J. Loewenstein}, \bibinfo{person}{S. Kelly}, {and} \bibinfo{person}{D. Shire}.} \bibinfo{year}{2003}\natexlab{}.
\newblock \showarticletitle{Perceptual efficacy of electrical stimulation of human retina with a microelectrode array during short-term surgical trials}.
\newblock \bibinfo{journal}{\emph{Invest Ophthalmol Vis Sci}} \bibinfo{volume}{44}, \bibinfo{number}{12} (\bibinfo{date}{Dec.} \bibinfo{year}{2003}), \bibinfo{pages}{5362--9}.
\newblock
\showISSN{0146-0404 (Print) 0146-0404 (Linking)}


\bibitem[Sadeghi et~al\mbox{.}(2024)]%
        {sadeghi_benefits_2024}
\bibfield{author}{\bibinfo{person}{Roksana Sadeghi}, \bibinfo{person}{Arathy Kartha}, \bibinfo{person}{Michael~P. Barry}, \bibinfo{person}{Paul Gibson}, \bibinfo{person}{Avi Caspi}, \bibinfo{person}{Arup Roy}, \bibinfo{person}{Duane~R. Geruschat}, {and} \bibinfo{person}{Gislin Dagnelie}.} \bibinfo{year}{2024}\natexlab{}.
\newblock \showarticletitle{Benefits of thermal and distance-filtered imaging for wayfinding with prosthetic vision}.
\newblock \bibinfo{journal}{\emph{Scientific Reports}} \bibinfo{volume}{14}, \bibinfo{number}{1} (\bibinfo{date}{Jan.} \bibinfo{year}{2024}), \bibinfo{pages}{1313}.
\newblock
\showISSN{2045-2322}
\urldef\tempurl%
\url{https://doi.org/10.1038/s41598-024-51798-x}
\showDOI{\tempurl}
\newblock
\shownote{Number: 1 Publisher: Nature Publishing Group}.


\bibitem[Sanchez-Garcia et~al\mbox{.}(2019)]%
        {sanchez-garcia_indoor_2019}
\bibfield{author}{\bibinfo{person}{Melani Sanchez-Garcia}, \bibinfo{person}{Ruben Martinez-Cantin}, {and} \bibinfo{person}{Josechu~J. Guerrero}.} \bibinfo{year}{2019}\natexlab{}.
\newblock \showarticletitle{Indoor {Scenes} {Understanding} for {Visual} {Prosthesis} with {Fully} {Convolutional} {Networks}}. In \bibinfo{booktitle}{\emph{{VISIGRAPP}}}.
\newblock
\urldef\tempurl%
\url{https://doi.org/10.5220/0007257602180225}
\showDOI{\tempurl}


\bibitem[Sanchez-Garcia et~al\mbox{.}(2020)]%
        {sanchez-garcia_semantic_2020}
\bibfield{author}{\bibinfo{person}{Melani Sanchez-Garcia}, \bibinfo{person}{Ruben Martinez-Cantin}, {and} \bibinfo{person}{Jose~J. Guerrero}.} \bibinfo{year}{2020}\natexlab{}.
\newblock \showarticletitle{Semantic and structural image segmentation for prosthetic vision}.
\newblock \bibinfo{journal}{\emph{PLOS ONE}} \bibinfo{volume}{15}, \bibinfo{number}{1} (\bibinfo{date}{Jan.} \bibinfo{year}{2020}), \bibinfo{pages}{e0227677}.
\newblock
\showISSN{1932-6203}
\urldef\tempurl%
\url{https://doi.org/10.1371/journal.pone.0227677}
\showDOI{\tempurl}


\bibitem[Sight(2013)]%
        {second_sight_argus_2013}
\bibfield{author}{\bibinfo{person}{Second Sight}.} \bibinfo{year}{2013}\natexlab{}.
\newblock \bibinfo{booktitle}{\emph{Argus® {II} {Retinal} {Prosthesis} {System} {Surgeon} {Manual}}}.
\newblock Number 900029-001 Rev C. \bibinfo{publisher}{Second Sight Medical Products, Inc.}, \bibinfo{address}{Sylmar, CA}.
\newblock
\urldef\tempurl%
\url{https://www.accessdata.fda.gov/cdrh_docs/pdf11/h110002c.pdf}
\showURL{%
\tempurl}


\bibitem[Sinclair et~al\mbox{.}(2016)]%
        {sinclair_appearance_2016}
\bibfield{author}{\bibinfo{person}{Nicholas~C. Sinclair}, \bibinfo{person}{Mohit~N. Shivdasani}, \bibinfo{person}{Thushara Perera}, \bibinfo{person}{Lisa~N. Gillespie}, \bibinfo{person}{Hugh~J. McDermott}, \bibinfo{person}{Lauren~N. Ayton}, {and} \bibinfo{person}{Peter~J. Blamey}.} \bibinfo{year}{2016}\natexlab{}.
\newblock \showarticletitle{The {Appearance} of {Phosphenes} {Elicited} {Using} a {Suprachoroidal} {Retinal} {Prosthesis}}.
\newblock \bibinfo{journal}{\emph{Investigative Ophthalmology \& Visual Science}} \bibinfo{volume}{57}, \bibinfo{number}{11} (\bibinfo{date}{Sept.} \bibinfo{year}{2016}), \bibinfo{pages}{4948--4961}.
\newblock
\showISSN{1552-5783}
\urldef\tempurl%
\url{https://doi.org/10.1167/iovs.15-18991}
\showDOI{\tempurl}


\bibitem[Stingl et~al\mbox{.}(2015)]%
        {stingl_subretinal_2015}
\bibfield{author}{\bibinfo{person}{K. Stingl}, \bibinfo{person}{K.~U. Bartz-Schmidt}, \bibinfo{person}{D. Besch}, \bibinfo{person}{C.~K. Chee}, \bibinfo{person}{C.~L. Cottriall}, \bibinfo{person}{F. Gekeler}, \bibinfo{person}{M. Groppe}, \bibinfo{person}{T.~L. Jackson}, \bibinfo{person}{R.~E. MacLaren}, \bibinfo{person}{A. Koitschev}, \bibinfo{person}{A. Kusnyerik}, \bibinfo{person}{J. Neffendorf}, \bibinfo{person}{J. Nemeth}, \bibinfo{person}{M.~A. Naeem}, \bibinfo{person}{T. Peters}, \bibinfo{person}{J.~D. Ramsden}, \bibinfo{person}{H. Sachs}, \bibinfo{person}{A. Simpson}, \bibinfo{person}{M.~S. Singh}, \bibinfo{person}{B. Wilhelm}, \bibinfo{person}{D. Wong}, {and} \bibinfo{person}{E. Zrenner}.} \bibinfo{year}{2015}\natexlab{}.
\newblock \showarticletitle{Subretinal {Visual} {Implant} {Alpha} {IMS}--{Clinical} trial interim report}.
\newblock \bibinfo{journal}{\emph{Vision Research}} \bibinfo{volume}{111}, \bibinfo{number}{Pt B} (\bibinfo{date}{June} \bibinfo{year}{2015}), \bibinfo{pages}{149--60}.
\newblock
\showISSN{1878-5646 (Electronic) 0042-6989 (Linking)}
\urldef\tempurl%
\url{https://doi.org/10.1016/j.visres.2015.03.001}
\showDOI{\tempurl}


\bibitem[Stronks and Dagnelie(2014)]%
        {stronks_functional_2014}
\bibfield{author}{\bibinfo{person}{H~Christiaan Stronks} {and} \bibinfo{person}{Gislin Dagnelie}.} \bibinfo{year}{2014}\natexlab{}.
\newblock \showarticletitle{The functional performance of the {Argus} {II} retinal prosthesis}.
\newblock \bibinfo{journal}{\emph{Expert Review of Medical Devices}} \bibinfo{volume}{11}, \bibinfo{number}{1} (\bibinfo{date}{Jan.} \bibinfo{year}{2014}), \bibinfo{pages}{23--30}.
\newblock
\showISSN{1743-4440}
\urldef\tempurl%
\url{https://doi.org/10.1586/17434440.2014.862494}
\showDOI{\tempurl}
\newblock
\shownote{Publisher: Taylor \& Francis \_eprint: https://doi.org/10.1586/17434440.2014.862494}.


\bibitem[Thorn et~al\mbox{.}(2021)]%
        {thorn_virtual_2021}
\bibfield{author}{\bibinfo{person}{Jacob~Thomas Thorn}, \bibinfo{person}{Naig Aurelia~Ludmilla Chenais}, \bibinfo{person}{Sandrine Hinrichs}, \bibinfo{person}{Marion Chatelain}, {and} \bibinfo{person}{Diego Ghezzi}.} \bibinfo{year}{2021}\natexlab{}.
\newblock \bibinfo{booktitle}{\emph{Virtual reality validation of naturalistic modulation strategies to counteract fading in retinal stimulation}}.
\newblock \bibinfo{type}{{T}echnical {R}eport}. \bibinfo{pages}{2021.11.17.468930} pages.
\newblock
\urldef\tempurl%
\url{https://doi.org/10.1101/2021.11.17.468930}
\showDOI{\tempurl}
\newblock
\shownote{Company: Cold Spring Harbor Laboratory Distributor: Cold Spring Harbor Laboratory Label: Cold Spring Harbor Laboratory Section: New Results Type: article}.


\bibitem[Thorn et~al\mbox{.}(2020)]%
        {thorn_virtual_2020}
\bibfield{author}{\bibinfo{person}{Jacob~Thomas Thorn}, \bibinfo{person}{Enrico Migliorini}, {and} \bibinfo{person}{Diego Ghezzi}.} \bibinfo{year}{2020}\natexlab{}.
\newblock \showarticletitle{Virtual reality simulation of epiretinal stimulation highlights the relevance of the visual angle in prosthetic vision}.
\newblock \bibinfo{journal}{\emph{Journal of Neural Engineering}} \bibinfo{volume}{17}, \bibinfo{number}{5} (\bibinfo{date}{Nov.} \bibinfo{year}{2020}), \bibinfo{pages}{056019}.
\newblock
\showISSN{1741-2552}
\urldef\tempurl%
\url{https://doi.org/10.1088/1741-2552/abb5bc}
\showDOI{\tempurl}
\newblock
\shownote{Publisher: IOP Publishing}.


\bibitem[Titchener et~al\mbox{.}(2022)]%
        {titchener_second-generation_2022}
\bibfield{author}{\bibinfo{person}{Samuel~A. Titchener}, \bibinfo{person}{David A.~X. Nayagam}, \bibinfo{person}{Jessica Kvansakul}, \bibinfo{person}{Maria Kolic}, \bibinfo{person}{Elizabeth~K. Baglin}, \bibinfo{person}{Carla~J. Abbott}, \bibinfo{person}{Myra~B. McGuinness}, \bibinfo{person}{Lauren~N. Ayton}, \bibinfo{person}{Chi~D. Luu}, \bibinfo{person}{Steven Greenstein}, \bibinfo{person}{William~G. Kentler}, \bibinfo{person}{Mohit~N. Shivdasani}, \bibinfo{person}{Penelope~J. Allen}, {and} \bibinfo{person}{Matthew~A. Petoe}.} \bibinfo{year}{2022}\natexlab{}.
\newblock \showarticletitle{A {Second}-{Generation} (44-{Channel}) {Suprachoroidal} {Retinal} {Prosthesis}: {Long}-{Term} {Observation} of the {Electrode}–{Tissue} {Interface}}.
\newblock \bibinfo{journal}{\emph{Translational Vision Science \& Technology}} \bibinfo{volume}{11}, \bibinfo{number}{6} (\bibinfo{date}{June} \bibinfo{year}{2022}), \bibinfo{pages}{12}.
\newblock
\showISSN{2164-2591}
\urldef\tempurl%
\url{https://doi.org/10.1167/tvst.11.6.12}
\showDOI{\tempurl}


\bibitem[Troyk(2017)]%
        {troyk_intracortical_2017}
\bibfield{author}{\bibinfo{person}{Philip~R. Troyk}.} \bibinfo{year}{2017}\natexlab{}.
\newblock \showarticletitle{The {Intracortical} {Visual} {Prosthesis} {Project}}.
\newblock In \bibinfo{booktitle}{\emph{Artificial {Vision}: {A} {Clinical} {Guide}}}, \bibfield{editor}{\bibinfo{person}{Veit~Peter Gabel}} (Ed.). \bibinfo{publisher}{Springer International Publishing}, \bibinfo{address}{Cham}, \bibinfo{pages}{203--214}.
\newblock
\showISBNx{978-3-319-41876-6}
\urldef\tempurl%
\url{https://doi.org/10.1007/978-3-319-41876-6_16}
\showDOI{\tempurl}


\bibitem[Vergnieux et~al\mbox{.}(2017)]%
        {vergnieux_simplification_2017}
\bibfield{author}{\bibinfo{person}{Victor Vergnieux}, \bibinfo{person}{Marc J.-M. Macé}, {and} \bibinfo{person}{Christophe Jouffrais}.} \bibinfo{year}{2017}\natexlab{}.
\newblock \showarticletitle{Simplification of {Visual} {Rendering} in {Simulated} {Prosthetic} {Vision} {Facilitates} {Navigation}}.
\newblock \bibinfo{journal}{\emph{Artificial Organs}} \bibinfo{volume}{41}, \bibinfo{number}{9} (\bibinfo{date}{Sept.} \bibinfo{year}{2017}), \bibinfo{pages}{852--861}.
\newblock
\showISSN{0160-564X}
\urldef\tempurl%
\url{https://doi.org/10.1111/aor.12868}
\showDOI{\tempurl}
\newblock
\shownote{Publisher: John Wiley \& Sons, Ltd}.


\bibitem[Weiland et~al\mbox{.}(2016)]%
        {weiland_electrical_2016}
\bibfield{author}{\bibinfo{person}{James~D. Weiland}, \bibinfo{person}{Steven~T. Walston}, {and} \bibinfo{person}{Mark~S. Humayun}.} \bibinfo{year}{2016}\natexlab{}.
\newblock \showarticletitle{Electrical {Stimulation} of the {Retina} to {Produce} {Artificial} {Vision}}.
\newblock \bibinfo{journal}{\emph{Annual Review of Vision Science}} \bibinfo{volume}{2}, \bibinfo{number}{1} (\bibinfo{year}{2016}), \bibinfo{pages}{273--294}.
\newblock
\urldef\tempurl%
\url{https://doi.org/10.1146/annurev-vision-111815-114425}
\showDOI{\tempurl}


\bibitem[Wilke et~al\mbox{.}(2011)]%
        {wilke_electric_2011}
\bibfield{author}{\bibinfo{person}{R.~G.~H. Wilke}, \bibinfo{person}{G.~Khalili Moghadam}, \bibinfo{person}{N.~H. Lovell}, \bibinfo{person}{G.~J. Suaning}, {and} \bibinfo{person}{S. Dokos}.} \bibinfo{year}{2011}\natexlab{}.
\newblock \showarticletitle{Electric crosstalk impairs spatial resolution of multi-electrode arrays in retinal implants}.
\newblock \bibinfo{journal}{\emph{Journal of Neural Engineering}} \bibinfo{volume}{8}, \bibinfo{number}{4} (\bibinfo{date}{June} \bibinfo{year}{2011}), \bibinfo{pages}{046016}.
\newblock
\showISSN{1741-2552}
\urldef\tempurl%
\url{https://doi.org/10.1088/1741-2560/8/4/046016}
\showDOI{\tempurl}


\end{thebibliography}

\clearpage
\appendix

\section{Eye Tracking Accuracy of the HTC Vive Pro}
\label{sec:app-eye-tracking}

To assess the precision of the HTC Vive Pro's built-in eye tracker, $n=30$ participants tracked a moving on-screen dot ($\sim 2.4^\circ$ visual angle) using their eyes. 
The dot moved randomly between four corners positioned halfway between the center and edges of the screen. 
It traversed the distance between points over $2.5 \pm 0.5$ seconds and remained stationary at each location for 1.5 seconds.

The angular error, defined as the distance between the dot's center and the user's gaze location, was measured every \SI{0.1}{\second}. 
Measurements were taken during both fixation (when the dot was stationary) and pursuit (when it was moving). 
Mean angular error was \SI{1.904 \pm 2.048}{\degree} during fixation and \SI{1.838 \pm 1.660}{\degree} during pursuit, with no significant difference between the two conditions (t-test for non-equal variances, $p > 0.27$).

\begin{figure}[h!]
    \centering
    \includegraphics[width=\linewidth]{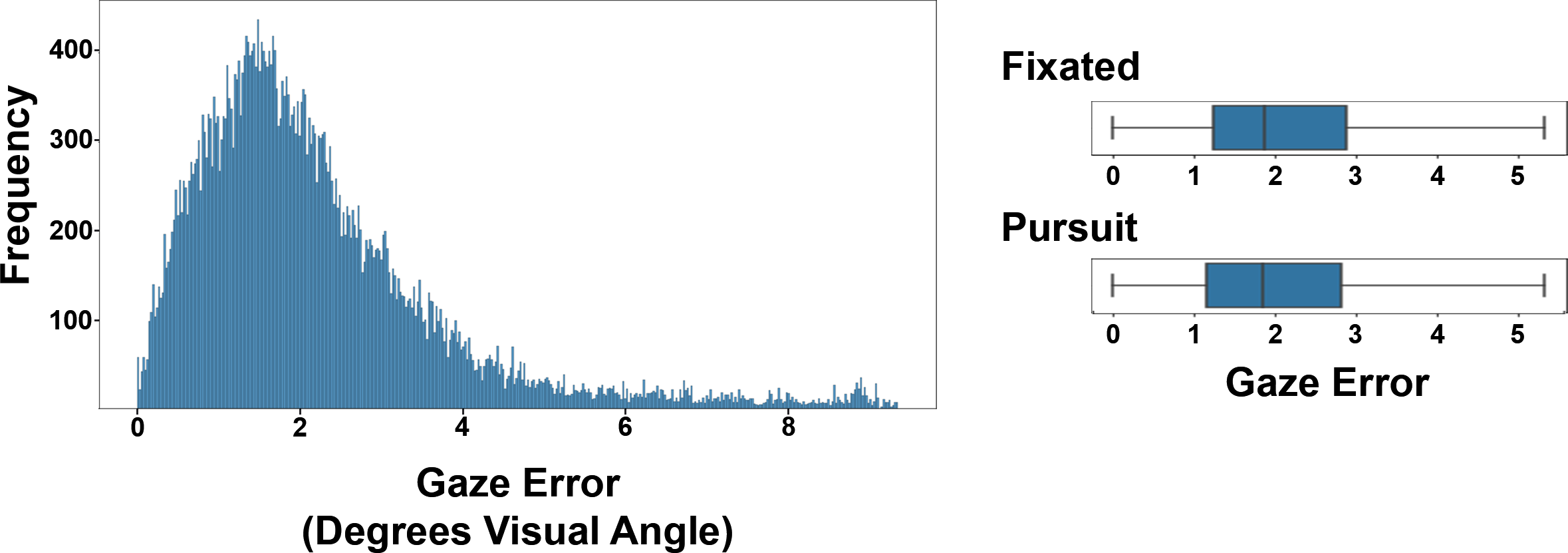}
    \caption{
        \textbf{Eye tracking accuracy of the HTC Vive Pro.} 
        The histogram (left) shows the distribution of angular gaze error (degrees visual angle) across all measurements, with most errors falling below \SI{5}{\degree}. 
        The boxplots (right) compare gaze error during fixation (when the dot was stationary) and pursuit (when the dot was moving). 
        Mean errors were similar across conditions (\SI{1.904 \pm 2.048}{\degree} for fixation and \SI{1.838 \pm 1.660}{\degree} for pursuit), with no significant difference between the two (t-test, $p > 0.27$). 
        Over 94\% of measurements had an error below \SI{5}{\degree}, confirming the system's precision for gaze-contingent rendering.
    }
    \label{fig:eye-tracking}
\end{figure}

Overall, 94.1\% of measurements had an angular error below \SI{5}{\degree}, and 80\% were below \SI{3}{\degree}. 
These results indicate that the HTC Vive Pro provides adequate precision for gaze-contingent rendering in simulated prosthetic vision experiments (Figure~\ref{fig:eye-tracking}).

\end{document}